\documentclass[prd ,aps, reprint, 11pt,nofootinbib,superscriptaddress,preprintnumbers]{revtex4-2}
\pdfoutput=1
\usepackage{geometry}
\usepackage{graphicx}
\usepackage{amssymb}
\usepackage{mathtools}
\usepackage{epstopdf}
\usepackage{color}
\usepackage{bbold}
\usepackage{fullpage}
\usepackage{float}
\usepackage[title]{appendix}
\usepackage[colorlinks, citecolor=blue, linkcolor=black, urlcolor=purple]{hyperref}
\usepackage{amsfonts}
\usepackage{amsmath}
\usepackage{setspace}
\usepackage{epsfig}
\usepackage[caption=false]{subfig}
\usepackage[usenames, dvipsnames]{xcolor}
\usepackage{tabularx}
\usepackage{empheq}
\usepackage{mathrsfs}
\usepackage{xcolor}

\usepackage{units} 
\newcommand{\lsim}{\lesssim}

\newcommand{\be}{\begin{eqnarray}}
\newcommand{\ee}{\end{eqnarray}}
\newcommand{\bea}{\begin{eqnarray}}
\newcommand{\eea}{\end{eqnarray}}
\newcommand{\nn}{\nonumber}

\newcommand{\beq}{\begin{equation}}
\newcommand{\eeq}{\end{equation}}

\newcommand{\vev}{\langle h \rangle}

\definecolor{darkblue}{rgb}{0.2,0.2,0.9}
\definecolor{colorRTD}{rgb}{.2,.2,.7}

\definecolor{colorHD}{rgb}{.2,0.9,.0.9}

\begin{document}

\title{The Weak Scale as a Trigger}

\author{Nima Arkani-Hamed}
\affiliation{School of Natural Sciences, Institute for Advanced Study, Princeton, New Jersey 08540, USA}
\author{Raffaele Tito D'Agnolo}
\affiliation{Institut de Physique Th\'eorique, Universit\'e Paris Saclay, CEA, F-91191 Gif-sur-Yvette, France}
\author{Hyung Do Kim}
\affiliation{Department of Physics and Astronomy and Center for Theoretical Physics,
Seoul National University, Seoul 08826, Korea}


\begin{abstract}
Does the value of the Higgs mass parameter affect the expectation value of local operators in the Standard Model? For essentially all local operators the answer to this question is ``no", and this is one of the avatars of the hierarchy problem: nothing is ``triggered" when the Higgs mass parameter crosses zero. In this letter, we explore settings in which Higgs mass parameters {\it can} act as a ``trigger" for some local operators ${\cal O}_T$. In the Standard Model, this happens for ${\cal O}_T  = {\rm Tr} (G \tilde G)$. We also introduce a ``type-0" two Higgs doublet model, with a $Z_4$ symmetry, for which ${\cal O}_T = H_1 H_2$ is triggered by the Higgs masses, demanding the existence of new Higgs states necessarily comparable to or lighter than the weak scale, with no wiggle room to decouple them whatsoever. Surprisingly, this model is not yet entirely excluded by collider searches, and will be incisively probed by the high-luminosity run of the LHC, as well as future Higgs factories. We also discuss a possibility for using this trigger to explain the origin of the weak scale, invoking a  landscape of extremely light, weakly interacting scalars $\phi_i$, with a coupling to ${\cal O}_T$ needed to make it possible to find vacua with small enough cosmological constant. The weak scale trigger links the tuning of the Higgs mass to that of the cosmological constant, while coherent oscillations of the $\phi_i$ can constitute dark matter.

\end{abstract}

\maketitle
\onecolumngrid

\tableofcontents

\section{Introduction} 

The apparent failure of naturalness in accounting for the minuscule size of both the cosmological constant and the Higgs mass, is giving us a profound structural clue about the the laws of fundamental physics. One of many ways of describing the hierarchy problem is in terms of how physics depends on the mass parameter of the Higgs, $m_h^2$. Finding $m_h^2 \ll \Lambda_H^2$, where $\Lambda_H$ is a UV scale for the Standard Model effective field theory, is mysterious because there is nothing special about $m_h^2=0$ for scalars; there is no difference in the number of degrees of freedom for massless versus massive spin zero particles, nor any obvious difference in the number of symmetries when $m_h^2 = 0$. Thus most of the ``dynamical" approaches to the hierarchy problem embed the Higgs in a larger structure, where $m_h^2$ is tied to other parameters that {\it are}  associated with symmetry enhancements when $m_h^2 = 0$, be it in the context of supersymmetry (where the chiral symmetry of fermion superpartners protects scalar masses), or theories of the Higgs as a pseudo-goldstone boson in either their four-dimensional or AdS avatars (where approximate shift symmetries play this role).  

What {\it does} vary, as we change the Higgs mass parameter $m_h^2$? Obviously the spectrum of the Standard Model changes, and this is detected by the non-trivial $m_h^2$ dependence of the two-point function propagators of the gauge bosons, fermions and the Higgs. For instance the gauge-invariant electron two-point function, $\bar{e}_{\dot{\alpha}} (x) W(x,y) e_{\alpha}(y)$, where $W(x,y)$ is an appropriate Wilson line, depends on the distance between the two spacetime points $(x-y)$ and certainly does strongly depend on $m_h^2$.

But we can also ask if there is any gauge invariant {\it local} operators ${\cal O}(x)$, whose vacuum expectation value is sensitive to $m_h^2$. We can probe $\langle \cal O \rangle$ by coupling $\cal O$, parametrically weakly, to some scalar $\phi$ via the coupling $\xi \phi {\cal O}$, and looking at the effective action induced for $\phi$. At tree-level, obviously ${\cal O}_h = h^\dagger h$ depends on $m_h^2$. But of course, once loop corrections are taken into account, $\langle {\cal O}_h \rangle$ is {\it not calculable} in the SM, which is one of the aspects of the hierarchy problem. We can simply look at the tadpole diagram, 
from $\xi \phi h^\dagger h$ which induces $\xi \phi \Lambda_H^2$ where $\Lambda_H$ is the cutoff for the Higgs sector. This is completely insensitive to $m_h^2$, and indeed $\langle h^\dagger h \rangle$ is essentially independent of the magnitude or sign of $m_h^2$. 
Continuing this line of thought leads to a more invariant characterization of ``tuning" associated with solutions of the hierarchy problem. Recall that the hierarchy problem is sharply posed in theories that allow the Higgs mass squared to be calculable, rather than taken as an input parameter. A closely related characterization is to find a theory in which $\langle h^\dagger h \rangle$ is calculable. In supersymmetric theories, $\langle h^\dagger h \rangle \sim m_{\rm SUSY}^2$ ($m_{\rm SUSY}$ is the soft supersymmetry breaking mass) while in composite Higgs models, $\langle h^\dagger h \rangle \sim f_\pi^2$ ($f_\pi$ is the decay constant of the composite meson/pion). From this perspective, the `degree of tuning' becomes a well defined ratio $
r \sim \frac{m_h^2}{\langle h^\dagger h \rangle}$, and in all known theories where both $m_h^2$ and $\langle h^\dagger h \rangle$ are calculable, making $r$ tiny requires the usual fine-tuning of parameters in the ultra-violet (UV) theory.

In this paper, we explore a different line of attack on the hierarchy problem. We will look for operators $\cal O$ that {\it are} sensitive to, or {\it triggered by}, scalar $m^2$ parameters. In the SM itself there is essentially a unique option--${\cal O}_G = {\rm Tr} (G \tilde{G})$. Another simple possibility for ${\cal O}$ presents itself in a two-Higgs doublet extension of the Standard Model, with Higgses $H_1,H_2$. With the crucial imposition of a $Z_4$ symmetry under which the product $(H_1 H_2) \to - (H_1 H_2)$, the operator ${\cal O}_H = H_1 H_2$ is triggered by $m_{1,2}^2$. We dub this the ``type-0" Two-Higgs Doublet Model (2HDM). 

\begin{figure*}[!t]
\begin{center}
\includegraphics[width=\textwidth]{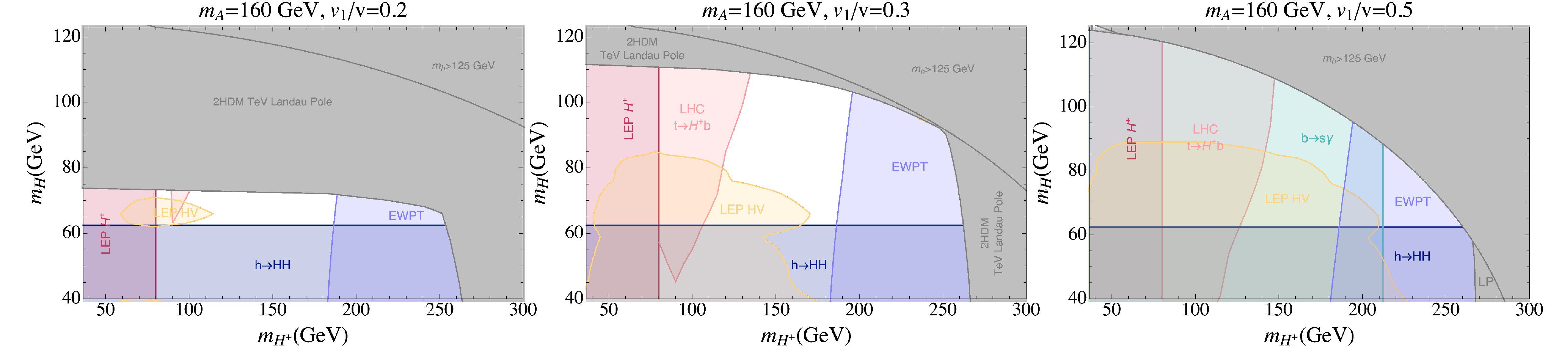}
\end{center}
\vspace{-.3cm}
\caption{Experimental Constraints on the type-0 2HDM in the $m_{H^\pm}-m_{H}$ plane (masses of the charged and new CP-even Higgses). The three panels correspond to three different choices for the vev of the new Higgs doublet $v_1=\langle H_1^0 \rangle=(0.2,0.3,0.5)v$. The mass of the new CP-Odd Higgs $m_A$ is fixed to 160 GeV. In all three cases we allow a 10$\%$ tuning of quartics, i.e. we take $\lambda_3+\lambda_4+\lambda_5=0.1 |\lambda_4+\lambda_5|=0.2 (m_{H^\pm}/v)^2$. In red we show the bound from $H^\pm$ pair production at LEP~\cite{Abbiendi:2013hk, ALEPH:2002aa} and in yellow from $HZ$ associated production and decays to fermions~\cite{Schael:2006cr}. In light blue we display the bound from ElectroWeak Precision Tests~\cite{ALEPH:2005ab} on the $S$, $T$ and $U$ oblique parameters~\cite{Peskin:1991sw,Altarelli:1990zd}. In light green we show bounds from searches for $B\to X_s \gamma$~\cite{Amhis:2019ckw,Arbey:2017gmh}. Indirect constraints from Higgs coupling measurements set an upper bound on ${\rm BR}(h\to HH)$ (in blue). The impact of LHC searches for $t\to H^+ b$ is shown in pink~\cite{CMS:2016szv, CMS:2019sxh, CMS:2016qoa, CMS:2014kga}. Theoretical constraints (in gray) from low energy Landau poles and the SM Higgs mass are summarized at the beginning of Section~\ref{sec:exp}. At high masses there is no solution for the quartics which gives $m_h=125$~GeV.}
\label{fig:intro_pheno}
\end{figure*}

Demanding this $Z_4$ symmetry and requiring that ${\cal O}_H$ is triggered put very tight constraints on the parameters of the model.  Both Higgses are forced to get a vev, while the $Z_4$ forbids the $B\mu$ term $B\mu H_1 H_2$ and so the new Higgses can only be raised in mass via quartic couplings, demanding the existence of new Higgs states necessarily comparable to or lighter than the weak scale, with no wiggle room to decouple them whatsoever. This is a large departure from the standard picture of electroweak symmetry breaking with a single Higgs, but to our great surprise, it is not (yet) entirely excluded by collider searches! Some representative region of parameter space that is still viable, in the plane of new CP even and charged Higgs masses, is shown in Fig.~\ref{fig:intro_pheno}, and the collider bounds will be discussed at greater length in Section~\ref{sec:2HDM}. There is a reasonable (if admittedly moderately tuned) region of parameter space where the new Higgs states have thus far escaped detection. This model will incisively be probed in the high-luminosity run of the LHC, and the new states can also be copiously produced at Higgs factories. If these new light states are seen, and the associated fingerprint of the $Z_4$ symmetry is confirmed, that would give direct experimental evidence for the ``weak scale as a trigger".

Given some operator ${\cal O}$ triggered by the weak scale, it is natural to try to use this trigger to attack the hierarchy problem in a new way. For instance, we can look for cosmological vacuum selection scenarios, that force
\bea
\mu^2 = \langle {\cal O}_H \rangle,
\eea
in the range
\bea
\mu_S^2 \ll \mu^2 \ll \mu_B^2,
\eea
Using ${\cal O}_H = H_1 H_2$ of our type-0 2HDM, this would force tuning for light Higgses. We present one such vacuum selection mechanism in the context of the landscape\footnote{A different approach to doing this with similar motivations was pursued in~\cite{Arvanitaki:2016xds}.}. We give a field-theoretic model for the landscape, with a ``UV landscape" containing moderately many vacua, not enough to find vacua with our small cosmological constant (CC). But we also imagine a separate ``IR landscape", with $n_\phi$ ultra-light, weakly coupled scalars $\phi_i$, each with a (spontaneously broken) $Z_2$ discrete symmetry potentially giving a factor of $2^{n_\phi}$ more vacua.  The $\phi_i$ also couple to ${\cal O}_H$. If $\langle {\cal O}_H \rangle$ is too small, the $2^{n_\phi}$ vacua of the $\phi_i$ sector are all degenerate and they don't help with making smaller vacuum energies possible. If $\langle {\cal O}_H \rangle$ is too big, the symmetry is broken so badly that only one vacuum remains for each $\phi_i$, and there is again no way to find small vacuum energy. The only way to find small vacuum energy is to tune Higgs vacuum expectation values so that $\mu_S^2 < \langle {\cal O}_H \rangle < \mu_B^2$. Thus using the weak scale as a trigger allows us to tie solutions to the cosmological constant and hierarchy problems. In this scenario, at least part of the landscape is associated with ultralight fields, which are not frozen to their minima but are oscillating in the universe today, providing novel possibilities for coherently oscillating dark matter, and mediating (extremely weak) long-range forces.  

\section{The Weak Scale as a Trigger}\label{sec:trigger}

As with the example of the operator $h^\dagger h$ discussed above,  almost all gauge invariant local operators in the SM have UV sensitive expectation values and are thus independent of $m_h^2$. Consider for example a Yukawa coupling, ${\cal O}_q = q h u^c$. If we add to the Lagrangian $\xi_q \phi {\cal O}_q$ to probe its vev,  at two-loops we generate 
\begin{figure*}[!t]
\begin{center}
\includegraphics[width=0.4\textwidth]{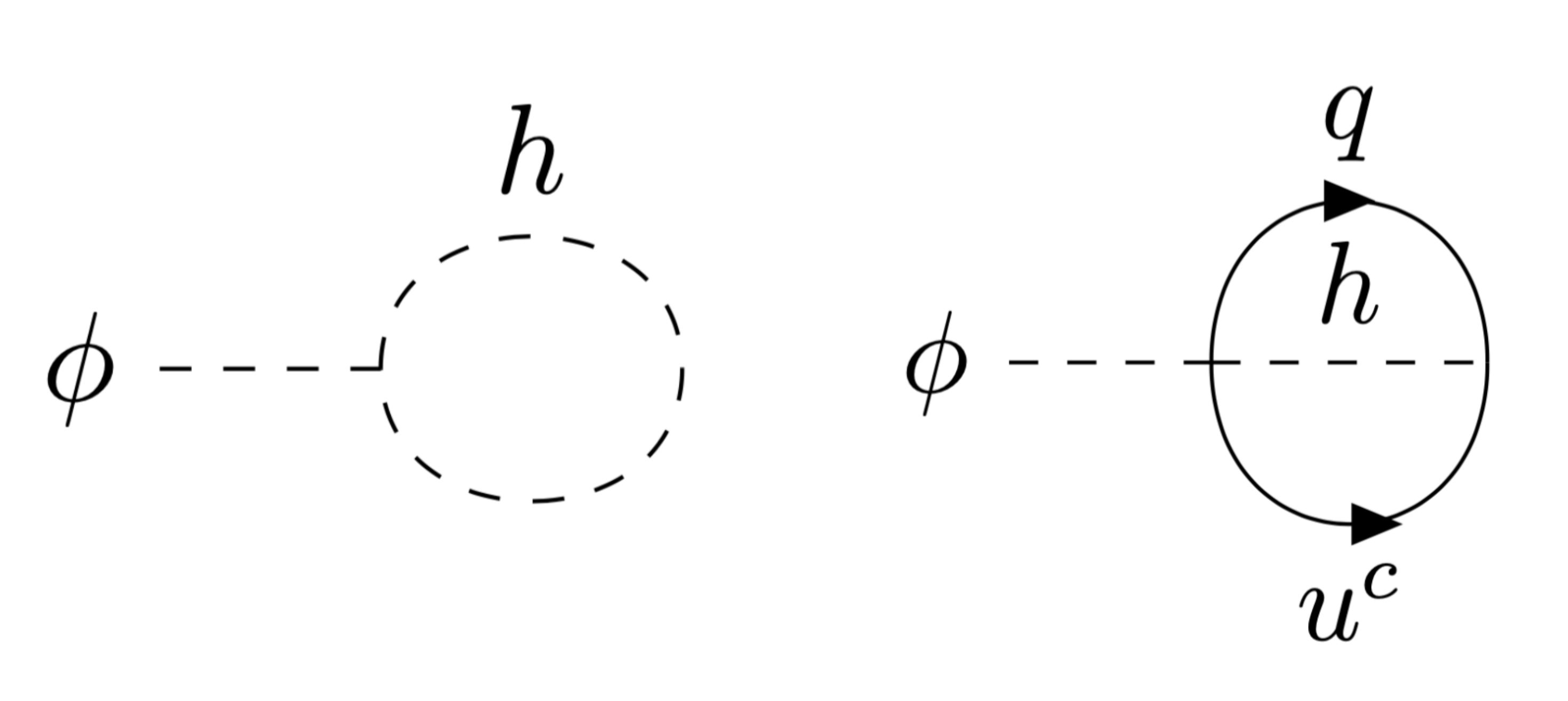}
\end{center}
\vspace{-.3cm}
\caption{Almost all gauge invariant local operators $\mathcal{O}$ in the SM have UV sensitive expectation values that are thus independent of $m_h^2$. We can probe their vevs by adding to the Lagrangian the parametrically weak interaction $\phi \mathcal{O}$ and look at the effective action for $\phi$. If we add, for example, $\phi |h|^2$ or $\phi q h u^c$ we can always close the loops in this Figure and obtain a tadpole for $\phi$ proportional to the cutoff. This happens because there are no global symmetries carried by $\cal O$, that are not already broken by the presence of ${\cal O}$ in the SM effective Lagrangian.}
\label{fig:tadpole}
\end{figure*}
the tadpole in Fig.~\ref{fig:tadpole} $\sim \frac{\xi_q y_q}{(16\pi^2)^2} \phi \Lambda^4$, proportional to the cutoff $\Lambda$. The reason is that every gauge invariant local operator already appears in the SM effective Lagrangian and we can close the loops. Said more invariantly, there are no global symmetries carried by relevant or marginal operators $\cal O$ in the SM, that are not broken by the presence of ${\cal O}$ in the effective Lagrangian. Thus if we have $\xi \phi \mathcal{O}$ in the Lagrangian, $\phi$ is not charged under any symmetries. So nothing forbids $\xi \phi \Lambda^n$ in the effective action. We can consider operators that are charged under the accidental baryon and lepton number global symmetries of the Standard Model; now these expectation values are not UV sensitive, but the global symmetries ensure that the expectation values for  these operators are equal to zero, again independent of the value of $m_h^2$. 
We can also imagine that $\phi$ has a shift symmetry $\phi \to \phi + c$, or equivalently, ask that ${\cal O} = \partial_{\mu} V^{\mu}$ is a total derivative. Then, $\phi {\cal O} = \phi \partial_{\mu} V^{\mu} = -(\partial_{\mu}\phi) V^{\mu}$. Again the expectation value of $\cal O$ is indeed UV insensitive, but at the same time $\langle {\cal O} \rangle =0$ independently of $m_h^2$.

There is one famous loophole to this argument, associated with the operator $G \tilde{G}$, which is the total derivative of a non-gauge invariant current, $G \tilde{G} = \partial_{\mu} K^{\mu}$, and can be turned on instanton backgrounds which break the shift symmetry non-perturbatively. Of course $G \tilde{G}$ is parity odd, but $\langle G \tilde{G} \rangle$ can be non-zero with a $\theta$ term. In pure QCD, we have $\langle G \tilde{G} \rangle \sim \theta \Lambda_{\rm QCD}^4$. With light quarks, we can rotate $(\theta + \xi \phi) G \tilde{G}$ into the quark mass matrix with an anomalous chiral rotation and use chiral perturbation theory to get the effective action $\sim m_{\pi}^2 f_{\pi}^2 \cos (\theta + \xi \phi) \sim m_{\pi}^2 f_{\pi}^2 + \theta \xi \phi m_{\pi}^2 f_{\pi}^2$ so $ \langle G \tilde{G} \rangle \sim \theta (m_u + m_d) \Lambda_{\rm QCD}^3$. This is suppressed by $\theta$ but is triggered by the weak scale through $(m_u+m_d)$. This triggering of $G \tilde{G}$ is used in the cosmological dynamics of relaxion models~\cite{Graham:2015cka}. We can also use triggered $\langle G \tilde{G} \rangle$ in conjunction with our model of low energy landscapes. We defer that discussion to Section~\ref{sec:SMtrigger}. First we discuss a simpler example of the weak scale as a trigger in 2HDM extensions of the Higgs sector.

\subsection{The Weak Scale as a Trigger in the type-0 2HDM}

We now consider a two-Higgs doublet extension of the Standard Model, with Higgs scalars $H_1,H_2$. Here the operator ${\cal O}_H =H_1 H_2$ is a good candidate to act as a trigger\footnote{This operator was already considered in the context of the relaxion in~\cite{Espinosa:2015eda} and in a different capacity related to fine-tuning in~\cite{Dvali:2001sm}.}. We want $(H_1 H_2)$ to be charged under a discrete symmetry which we can probe by coupling to some $\phi$ with $\xi \phi H_1 H_2$. The simplest choice is a symmetry under which $\phi \to -\phi$ and $(H_1 H_2) \to - (H_1 H_2)$. This is part of the $Z_4$ symmetry
\be
H_1 \to i e^{i\alpha} H_1, \quad H_2 \to i e^{-i\alpha} H_2, \quad \phi \to -\phi \label{eq:Z4_I}
\ee
with $\alpha$ in $U(1)_Y$. The $Z_4$ symmetry acts also on quark and lepton bilinears and we have $2^3$ possible charge assignments: $\pm i e^{-i\alpha}$ for $q u^c, q d^c, l e^c$. The case where a single Higgs couples to the quarks and leptons will be phenomenologically safest, so we will focus on that in the following. This fixes the fermion charge assignments, giving the $Z_4$ symmetry
\bea
H_1 \to +i e^{i\alpha} H_1, \quad 
H_2 &\to& +i e^{-i\alpha} H_2, \quad (H_1 H_2) \to - (H_1 H_2), \nn \\
(q u^c) \to -i e^{i\alpha} (q u^c), \quad
(q d^c) &\to& +i e^{-i\alpha} (q d^c), \quad
(l e^c) \to +i e^{-i\alpha} (l e^c). 
\eea
The renormalizable $H_{1,2}$ potential invariant under this symmetry is
\be
V&=&V_{H_1 H_2}+V_Y\, , \nn \\
V_{H_1 H_2}&=& m_1^2|H_1|^2+m_2^2|H_2|^2+\frac{\lambda_1}{2}|H_1|^4+\frac{\lambda_2}{2}|H_2|^4 \nn \\
&+&\lambda_3 |H_1|^2|H_2|^2 +\lambda_4 |H_1 H_2|^2 
+\left(\frac{\lambda_5}{2}(H_1 H_2)^2 +{\rm h.c.}\right)\, , \nn \\
V_Y&=& Y_u q H_2 u^c +Y_d q H_2^\dagger d^c + Y_e l H_2^\dagger  e^c\, .
\label{eq:2HDM_t0}
\ee
Note the absence of the $B\mu$-term, $B\mu H_1 H_2$ and of the two quartics $\lambda_{6,7}|H_{1,2}|^2(H_1H_2)$, all forbidden by the $Z_4$ symmetry. Note also the $\lambda_5 (H_1 H_2)^2$ term which is allowed. Without this term, the potential would have an accidental Peccei-Quinn (PQ) symmetry and would yield a weak scale axion. 

It is very important that $B\mu = \lambda_{6,7}=0$, otherwise we would have $m_{1,2}^2$-independent contributions to the vev of our trigger operator from Fig.~\ref{fig:tadpole_trigger}, as for instance
\be
\mu^2 \equiv \langle H_1 H_2 \rangle\sim \xi \phi B\mu \log \frac{\Lambda^2}{|m_H^2|}.
\ee
where for simplicity we have taken the Higgs masses to a common value $m_H^2$. On the contrary, if $B\mu=\lambda_{6,7}=0$, then $\mu^2 (m_1^2, m_2^2)$ is a UV-insensitive, calculable function of $m_1^2$, $m_2^2$ for which the weak scale is a trigger. This is a consequence of the $U(1)$ PQ symmetry of the potential in Eq.~\eqref{eq:2HDM_t0}. $H_1 H_2$ has charge 1 under this symmetry. The only explicit breaking of the PQ is by the quartic $\lambda_5 (H_1 H_2)^2$, for which $\lambda_5$ has charge $-2$, and so no analytic expression in the couplings can give something of charge 1. This is shown schematically in Fig.~\ref{fig:tadpole_trigger}.

\begin{figure*}[!t]
\begin{center}
\includegraphics[width=0.48\textwidth]{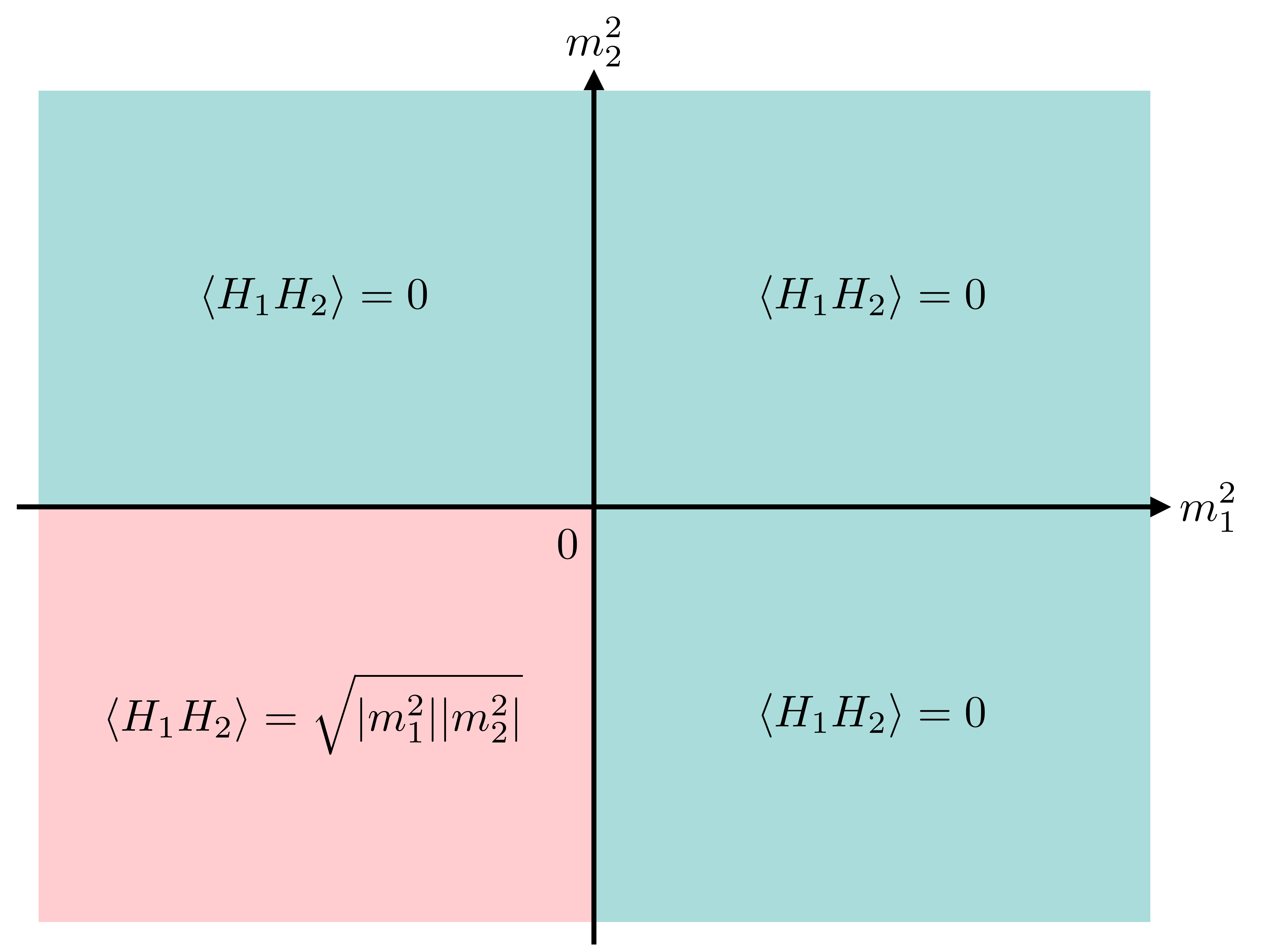}
\includegraphics[width=0.48\textwidth]{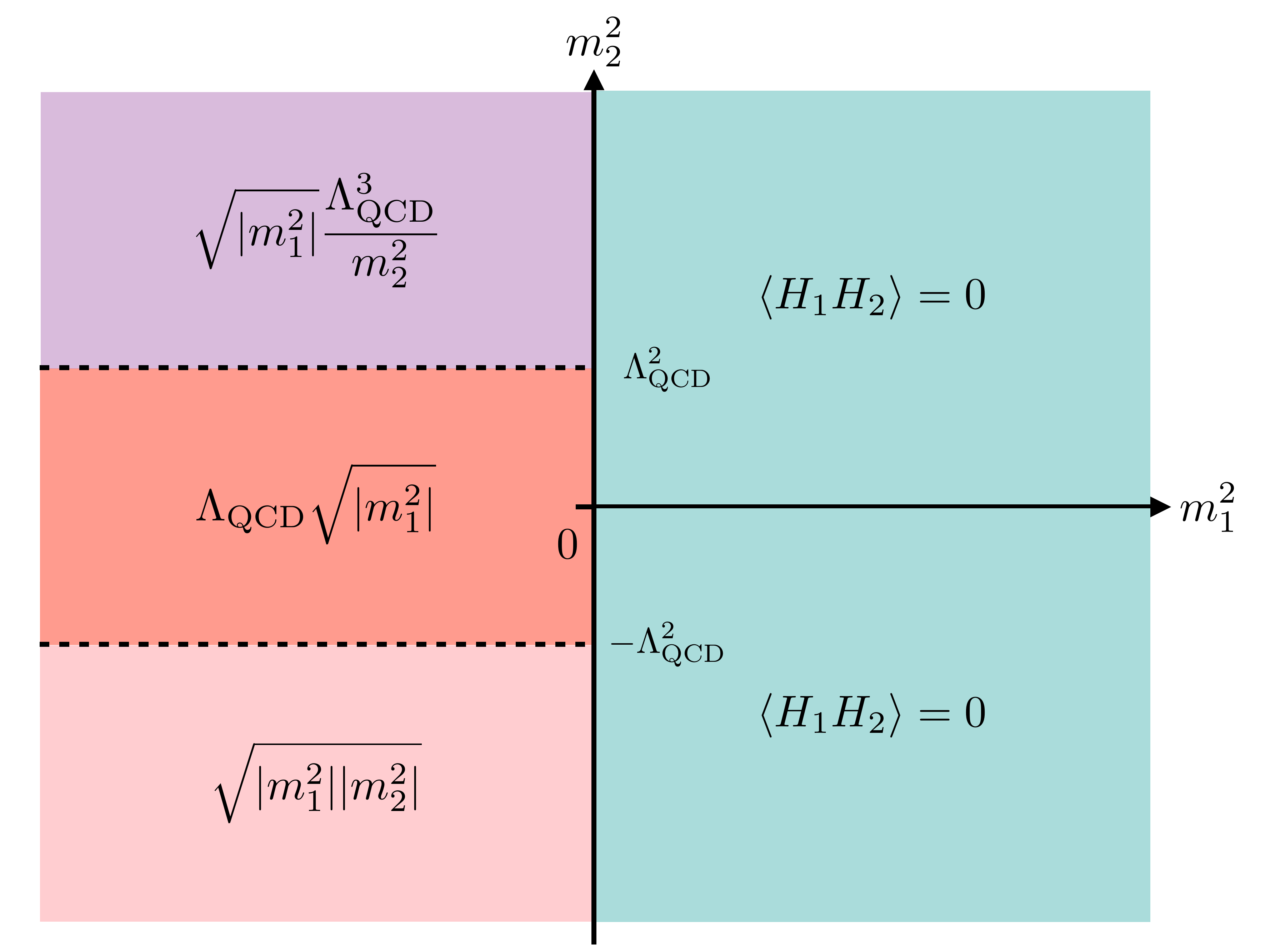}
\end{center}
\vspace{-.3cm}
\caption{In the type-0 2HDM (Eq.~\eqref{eq:2HDM_t0}), $\langle H_1 H_2 \rangle$ is a UV-insensitive, calculable function of the masses of the two Higgses: $m_1^2$, $m_2^2$. In the left panel we show its classical value while in the right one we include quantum effects. $m_1^2$, $m_2^2$ in the Figure are effective masses that include contributions from cross quartic couplings. $\Lambda_{\rm QCD}$ is the QCD scale with all quark masses below $\Lambda_{\rm QCD}$.}
\label{fig:vevs}
\end{figure*}

\begin{figure*}[!t]
\begin{center}
\includegraphics[width=0.18\textwidth]{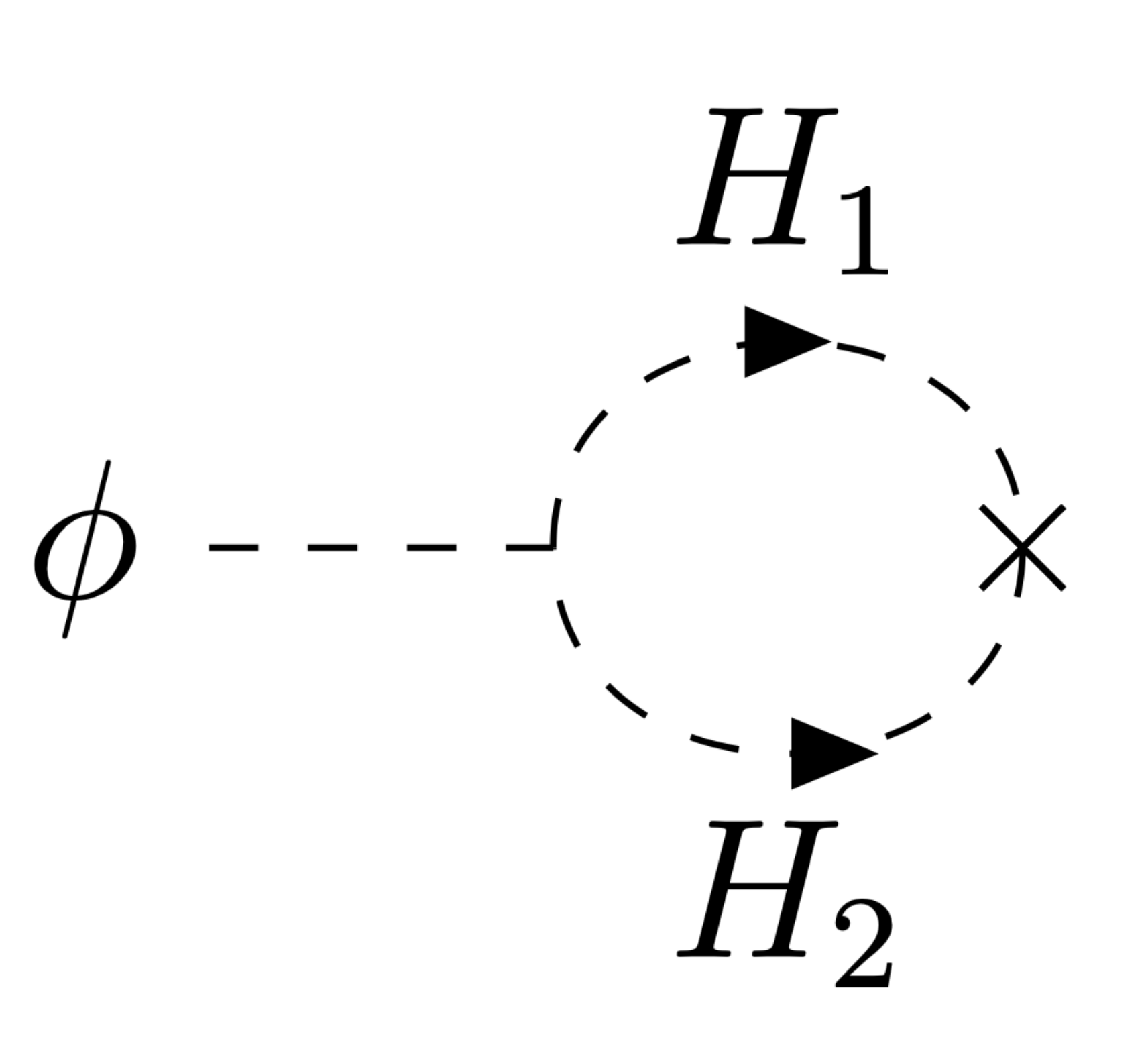}
\end{center}
\vspace{-.3cm}
\caption{In a 2HDM with a $Z_4$ symmetry (Eq.s~\eqref{eq:Z4_I} and \eqref{eq:2HDM_t0})  the $H_1 H_2$ vev is a UV-insensitive, calculable function of the two Higgs masses. This can be seen by adding to the Lagrangian the parametrically weak interaction $\phi H_1 H_2$. We can only close the loop in this Figure and generate a $\phi$ tadpole independent of $\langle H_1 H_2 \rangle$ with an insertion of $B\mu, \lambda_6$ or $\lambda_7$ which break the $Z_4$ symmetry and are thus absent from our 2HDM potential in Eq.~\eqref{eq:2HDM_t0}.} 
\label{fig:tadpole_trigger}
\end{figure*}

Let us now see what is the value of $\mu^2$ as a function of $m_1^2$ and $m_2^2$. At tree level $\mu^2=0$ unless both $m_1^2$ and $m_2^2$ are negative. If they are both negative, we have
\be
\mu^2 = \langle H_1 \rangle \langle H_2 \rangle \sim \sqrt{\frac{|m_1^2||m_2^2|}{\lambda_1\lambda_2}}
\ee
where we have ignored all cross quartic couplings. For simplicity we will call $\mu^2\sim \sqrt{|m_1^2||m_2^2|}$ for $\lambda_{1,2}$ not too tiny. We will keep this characterization even including cross quartics. In this case $m_{1,2}^2$ should be interpreted as $(m_{1,2}^{ {\rm eff}})^2$ which include the cross quartics contributions. So, at tree level, we have the possible $\mu^2$ shown in the left panel of Fig.~\ref{fig:vevs}. 

The picture is a little more interesting taking QCD chiral symmetry breaking into account. In the following $\Lambda_{\rm QCD}$ is the QCD scale with all quark masses below $\Lambda_{\rm QCD}$.
If $m_{1,2}^2 > 0$, then we do not break $SU(2) \times U(1)$ above the QCD scale. Since all the $SU(2) \times U(1)$ invariants of charge $-1$ under the $Z_4$ involve $H_1$, in the low energy theory after integrating out $H_{1,2}$, there can be no linear terms in $\phi$. So we have to have one or both of $m_{1,2}^2 < 0$. 

Consider first $m_1^2 > 0$ and $m_2^2 < 0$. For  $|m_2^2| \gg m_1^2$ we can first integrate out $H_2$. In the effective theory containing the neutral and charged components of $H_1=(h_1^0, h_1^+)^T$ and $\phi$, $SU(2)\times U(1)$ is broken, but a $Z_2$ subgroup of the $Z_4$, 
\be
H_1 \to - H_1, \quad 
H_2 &\to& H_2, \quad \phi \to - \phi, \nn \\
(q u^c) \to (q u^c), \quad
(q d^c) &\to& (q d^c), \quad
(l e^c) \to (l e^c). \label{eq:Z4}
\ee
is preserved. This symmetry is still not broken also after integrating out $(h_1^0, h_1^+)$, so again in the low energy theory there are no linear terms in $\phi$. 

Instead for $m_1^2 \gg |m_2^2|$, after integrating out $H_1$, we still have $SU(2) \times U(1)$ but there is no operator of charge $-1$ under the $Z_4$ (again since all these involve $H_1$). So again, no linear term in $\phi$ is generated. Thus, if $m_1^2 > 0$, for any $m_2^2$, we have $\mu^2 = 0$.

Now consider $m_1^2 <0$. If $m_2^2 < 0$, we have $\mu^2 \sim \sqrt{|m_1^2|}\max[\sqrt{|m_2^2|}, \Lambda_{\rm QCD}]$. If $m_2^2 > 0$ and $m_2^2\gg m_1^2, \Lambda_{\rm QCD}^2$, we first integrate out $H_2$ obtaining
\be
\frac{\xi}{m_2^2} \phi H_1^* q u^c+...
\ee
Hence after chiral symmetry breaking, we have $\mu^2 \sim \frac{\sqrt{|m_1^2|} \Lambda_{\rm QCD}^3}{m_2^2}$. If $|m_2^2| \ll m_1^2$, we can first integrate out $H_1$ giving
\be
\xi \phi \sqrt{|m_1^2|} h_2^0+...
\ee
Then if $m_2^2\gg\Lambda_{\rm QCD}^2$ we get the same result as before: $\mu^2 \sim \frac{\sqrt{|m_1^2|} \Lambda_{\rm QCD}^3}{m_2^2}$. If instead $|m_2^2|\ll \Lambda_{\rm QCD}^2$, the quartic term dominates and the $H_2$ VEV is just set by $\Lambda_{\rm QCD}$ from the potential
\bea
V(h_2^0) \simeq y_t \Lambda_{\rm QCD}^3 h_2^0 + \lambda (h_2^0)^4,
\eea
so we have $\langle h_2^0 \rangle \sim \frac{\Lambda_{\rm QCD}}{\lambda^{1/3}} \sim \Lambda_{\rm QCD}$ for a not too tiny quartic $\lambda$. This discussion is summarized in the right panel of Fig. \ref{fig:vevs}.
We have seen that in a 2HDM with a $Z_4$ symmetry--what we have called the type-0 2HDM--$\mathcal{O}_H=H_1 H_2$ is a good weak scale trigger. We defer to Section~\ref{sec:landscape} an explicit construction that uses $\mathcal{O}_H$ to tie the cosmological constant to the value of the Higgs mass. In the next Section we explore the collider constraints on the type-0 2HDM.

In this Section we have worked in the limit $B\mu= \lambda_{6,7} = 0$. It is clear from the previous discussion that any $B\mu \ll \mu_S^2$ and $\lambda_{6, 7}\ll \mu_S^2/M_*^2$ do not affect our conclusions or the phenomenology of the type-0 2HDM. However if we take $B\mu$ and $\lambda_{6,7}$ to be exactly zero, the model has a  $Z_2$ symmetry under which $H_1 \to -H_1$. To avoid a domain wall problem in the early Universe, we need to introduce a tiny breaking of this symmetry, $B\mu \gtrsim v^4/M_{\rm Pl}^2$. 

In our Universe, just below the critical temperature of the EW phase transition, domains of size $\sim 1/v$ with different signs of the $H_1$ vev are formed inside any Hubble volume. The walls separating these domains have an energy per unit area $\sigma \simeq v^3$. If locally the walls have curvature $1/R$ they will try to flatten due to the tension $\sigma$ that acts as an effective pressure $p_T \simeq \sigma/R$. Parametrically we can consider two extreme regimes: no coupling between the wall and the SM thermal bath or reflection of every SM particle by the wall.

In absence of interactions with the SM bath, the walls expand at the speed of light until we have one domain per Hubble patch. After this initial expansion the energy density in domain walls is $\rho_{\rm W} \simeq v^3 H$, redshifting as $1/a^2$. At a temperature $T_D\simeq v(v/M_{\rm Pl})^{1/2}\simeq {\rm keV}$ the walls dominate the energy density of the Universe. Interactions with the SM thermal bath slow down the expansion of the walls. If we ignore small couplings and assume that every SM particle is reflected with unit probability, we have a pressure $p_F \simeq v_{\rm W} T^4$ that slows down the expansion of the wall. Here $v_{\rm W}$ is the velocity of the wall. Balancing $p_F$ with $p_T$ we obtain a steady state solution with $v_{\rm W} \simeq v^{3/2}/(M_{\rm Pl} H^{1/2})$ and $R\simeq v^{3/2}/(M_{\rm Pl} H^{3/2})$. In this regime, the energy density of domain walls redshifts as $1/a^3$, but the initial energy density after one Hubble time at the phase transition is larger than that in the absence of friction. In the end we obtain the same temperature of domain walls domination as before ($T_D\simeq v(v/M_{\rm Pl})^{1/2}\simeq {\rm keV}$). Note that both with and without friction the walls make up a negligibly small fraction of the energy density of the Universe during BBN.

In conclusion to avoid domain walls domination we need to introduce an energy difference between the two vacua $\pm v_1$, for example by turning on $B\mu H_1 H_2$ in the Lagrangian. If we ask that the acceleration provided by this term $B\mu/v$ is larger than Hubble at the time of domination $H(T_D)\simeq v^3/M_{\rm Pl}^2$, we obtain $B\mu \gtrsim v^4/M_{\rm Pl}^2$. It is easy to show that the wall subsequently collapse in approximately on Hubble time $\sim 1/H(T_D)$.

Interestingly, when we explicitly use $H_1 H_2$ as a trigger in our landscape model of Section~\ref{sec:landscape}, the scalars in the landscape spontaneously break the $Z_2$ and automatically provide a large enough $B\mu$ term to avoid domain wall domination.



\section{Collider Phenomenology of the $H_1 H_2$ Trigger: the type-0 2HDM}\label{sec:2HDM} 

\subsection{Masses and Couplings}
In the previous Section we have described the conditions that make $H_1 H_2$ a good weak scale trigger. We need to impose a $Z_4$ symmetry which sets to zero $B\mu, \lambda_6$ and $\lambda_7$. This gives the potential in Eq.~\eqref{eq:2HDM_t0}. We also choose the $Z_4$ charge assignment which allows only $H_2$ to couple to the quarks and leptons: 
\be
V_Y=Y_u q H_2 u^c +Y_d q H_2^\dagger d^c + Y_e l H_2^\dagger  e^c\, . \label{eq:Yukawas}
\ee
As we show in the following, even in this case it is not possible to decouple collider signatures. Many of the facts about Higgs couplings in this Section are well-known in the 2HDM literature, but we repeat them to be self-contained. For reviews we refer the reader to~\cite{Gunion:2002zf, Djouadi:2005gj, Branco:2011iw}.

Hermiticity makes the potential in Eq.~\eqref{eq:2HDM_t0} CP-conserving. The only coupling that can have a phase is $\lambda_5$, but the rephasing $H_1\to H_1 e^{-i {\rm arg}(\lambda_5)/2}$ has no effect on the other terms in the Lagrangian, so there is no mass mixing between CP-even and CP-odd states. The masses of the Higgs bosons in $H_{1,2}$ are
\be
m_A^2 &=& - v^2 \lambda_5\, , \nn \\
m_{H^\pm}^2 &=& - v^2 \frac{\lambda_5+\lambda_4}{2} \nn \\
m_{h,H}^2 &=& \frac{1}{2}\left(\lambda_1 v_1^2+\lambda_2 v_2^2\pm\sqrt{\left(\lambda_2 v_2^2-\lambda_1 v_1^2\right)^2+4 v_1^2v_2^2\lambda_{345}^2}\right) \, . \label{eq:masses}
\ee
For convenience we have defined $\lambda_{345}\equiv\lambda_3+\lambda_4+\lambda_5$. This parameter sets the strength of the mixing between the two CP-even Higgses.  Limits of enhanced symmetry include: 1) when $\lambda_5=0$ a PQ symmetry acting on $H_1 H_2$ is only spontaneously broken and $A$ becomes a massless Goldstone boson, 2) when $\lambda_4=\lambda_5$ the potential acquires a $SU(2)$ custodial symmetry under which $\mathcal{H}=(H^+, i A, H^-)$ transforms as a triplet, hence $m_A=m_{H^\pm}$. 

\begin{figure*}[!t]
\begin{center}
\includegraphics[width=\textwidth]{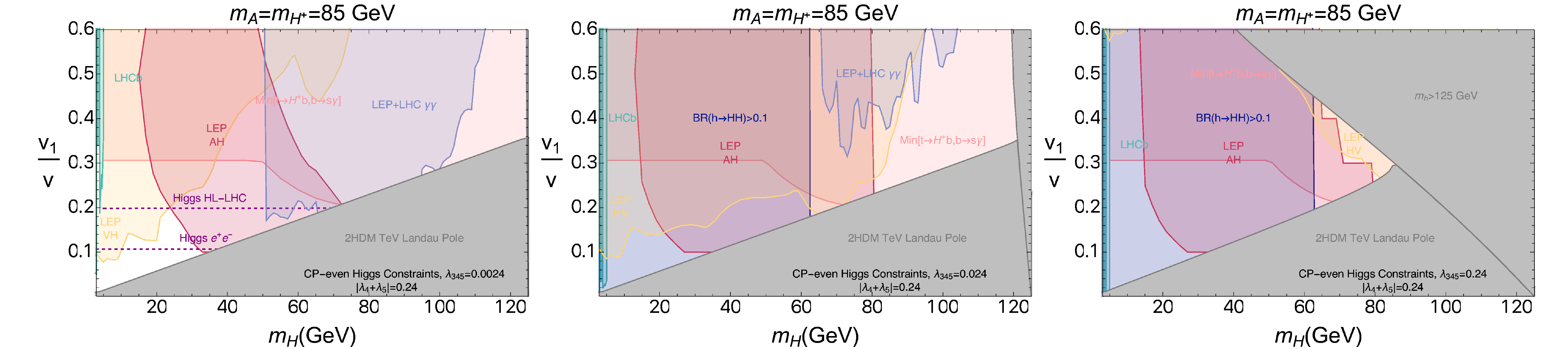} \\ \mbox{} \\
\includegraphics[width=\textwidth]{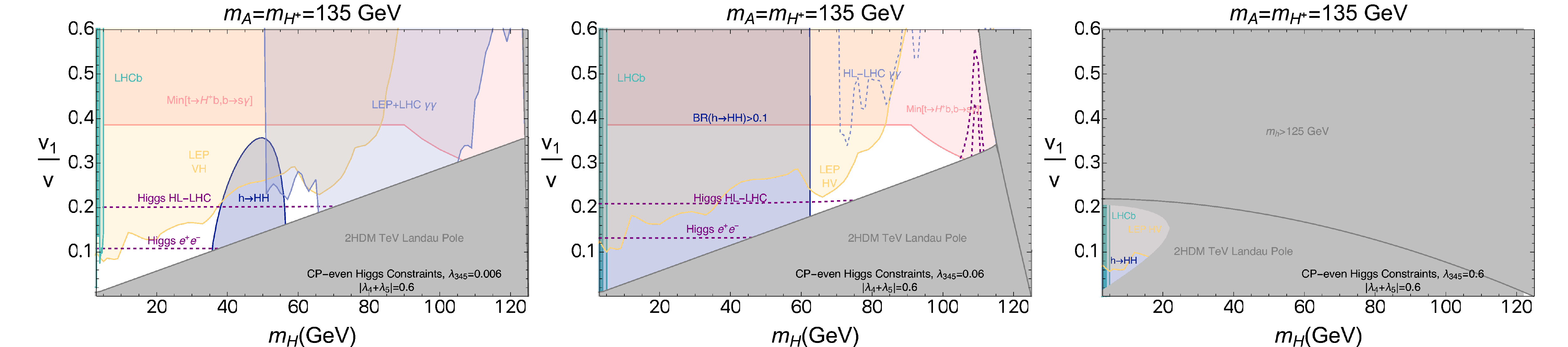} \\ \mbox{} \\
\includegraphics[width=\textwidth]{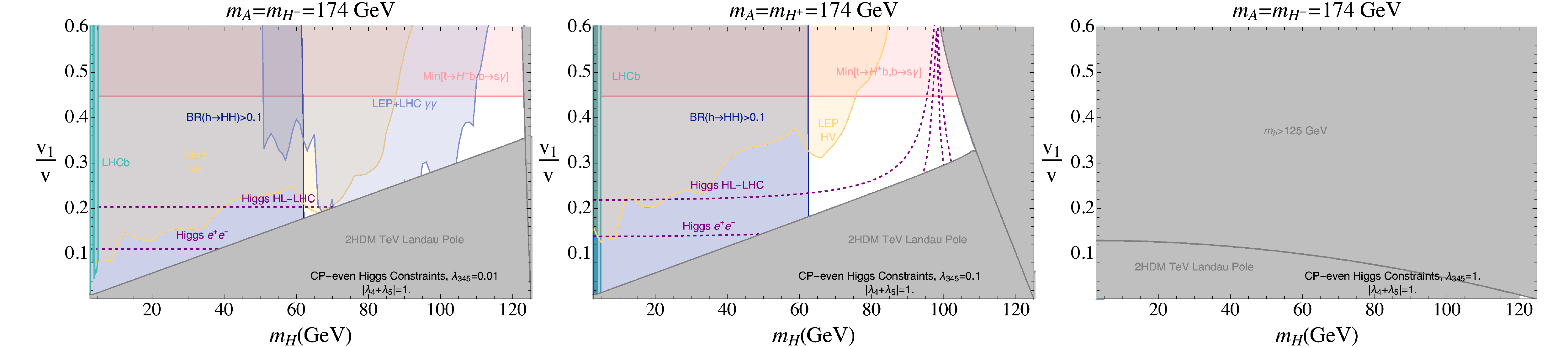} \\
\end{center}
\vspace{-.3cm}
\caption{Experimental Constraints on the CP-even Higgs $H$ for $m_{H^{\pm}}=m_A$ and different values of $\lambda_{345}$. From top to bottom we increase $m_{H^{\pm}}$. From left to right we move from $1\%$ tuning ($\lambda_{345}=0.01|\lambda_4+\lambda_5|$) to natural values of the quartics ($\lambda_{345}=|\lambda_4+\lambda_5|$).
In red we show the bound from $e^+e^-\to Z\to A H$ at LEP~\cite{Schael:2006cr} and in yellow from $HZ$ associated production~\cite{Schael:2006cr} followed by decays to fermions. In light blue we display the current sensitivity of $H\to \gamma\gamma$ at LEP and the LHC~\cite{ACHARD200228, Aad:2014ioa, Sirunyan:2018aui} and a projection for the HL-LHC obtained rescaling~\cite{Sirunyan:2018aui}. In light green we show bounds from searches for $B\to K^{(*)} H \to K^{(*)}\mu\mu$ at LHCb~\cite{Aaij:2012vr, Wei:2009zv}. Indirect constraints from Higgs coupling measurements (purple and blue) are discussed in Section~\ref{sec:indirect}. The pink shaded area shows the strongest bound point-by-point between searches for flavor changing processes, mainly $b\to s\gamma$~\cite{Amhis:2019ckw, Arbey:2017gmh}, and LHC searches for $t\to H^+ b$~\cite{CMS:2016szv, CMS:2019sxh, CMS:2016qoa, CMS:2014kga}. Theoretical constraints (in gray) from low energy Landau poles and the SM Higgs mass are summarized at the beginning of Section~\ref{sec:exp}.}
\label{fig:CPeven_Heavy}
\end{figure*}

Measurements of Higgs couplings and low energy flavor observables require $v_1 \lesssim v_2$, as shown in Fig.~\ref{fig:CPeven_Heavy}. In this limit the SM-like Higgs $h$ is heavier than its CP-even partner $H$. At leading order in $v_1/v$ we have
\be
m_h^2 &\simeq& \lambda_2 v^2 \nn \\
m_H^2 &\simeq& v_1^2 \left(\lambda_1 - \frac{\lambda_{345}^2}{\lambda_2}\right)\, .
\label{eq:massesII}
\ee
Eq.s~\eqref{eq:masses} and~\eqref{eq:massesII} are interesting from a phenomenological perspective, as they imply that the new Higgses all have masses comparable to $v$ or smaller. This is not surprising since we set to zero $B\mu$, the only other scale in the potential.  Before discussing laboratory constraints on these new particles it is useful to take a look at their couplings to the SM at leading order in $v_1/v$ (a more general approach to the decoupling limit was discussed in~\cite{Gunion:2002zf}). The charged and CP-odd Higgses have couplings to a pair of fermions suppressed by the small $H_1$ vev
\be
g_{H^+t^cb}\simeq g_{htt}^{\rm SM} \frac{v_1}{v}\, , \quad g_{H^-t b^c}\simeq g_{hbb}^{\rm SM} \frac{v_1}{v}\, , \quad
g_{A\psi\psi}\simeq \pm g_{h\psi\psi}^{\rm SM} \frac{v_1}{v} \, , \label{eq:charged_cpodd}
\ee
and no tree-level couplings to two SM gauge bosons, $g_{H^{\pm} W^{\mp}Z}\simeq 0\, , \; g_{H^{\pm} W^{\mp}\gamma}\simeq 0\, , \; g_{AVV}\simeq0$.
The coupling structure of the CP-even Higgs is slightly more complex. Its couplings also vanish in the small $v_1/v$ limit, but there is a second relevant parameter, $\lambda_{345}$. If we tune its value we can take either a fermiophobic or a bosophobic limit,
\be
g_{H\psi\psi}\simeq -g_{h\psi\psi}^{\rm SM} \frac{\lambda_{345}}{\lambda_2}\frac{v_1}{v}\, , \quad
g_{HVV}\simeq g_{hVV}^{\rm SM} \frac{|\lambda_2-\lambda_{345}|}{\lambda_2}\frac{v_1}{v}\, . \label{eq:Hcouplings}
\ee
$\lambda_{345}$ also controls the coupling between the SM Higgs and a pair of new Higgses
\be
\lambda_{h HH}\simeq \lambda_{345} v\, , \quad
\lambda_{h AA}\simeq (\lambda_{345}-2 \lambda_5) v\, , \label{eq:trilinear}
\ee
thus determining the decay width $h\to HH$. From Eq.s~\eqref{eq:charged_cpodd} and~\eqref{eq:Hcouplings} it is clear that the new states can not be easily decoupled. Decreasing $v_1$ reduces all couplings with a single new Higgs in the vertex, but makes $H$ lighter. Taking $\lambda_{345}\ll \lambda_2$ suppresses $H$ couplings to fermions, but maximizes those to gauge bosons. If we tune $\lambda_{345}\simeq \lambda_2$ we reduce couplings to gauge bosons, but those to fermions are unsuppressed.  Phenomenologically we find that small $\lambda_{345}$ is harder to detect, as shown in Fig.~\ref{fig:CPeven_Heavy}, but even in this limit $H$ is within reach of future colliders. Notice that we need at least $\lambda_5$ to be non-zero if we want $A$ and $H^\pm$ to be massive. In this case $\lambda_{345}=0$ does not give any extra symmetry, as can be seen for instance by inspecting one-loop RGEs~\cite{Branco:2011iw}. 

Pair production of the new states is completely fixed, insensitive to $v_1$ and other unknown parameters of the model. We do not list here $HHV$- and $HH VV$-type couplings, but they are $\mathcal{O}(g)$ and $\mathcal{O}(g^2)$, respectively, for all three new Higgses. They can be found for instance in~\cite{Djouadi:2005gj}. 

Now we have all the ingredients to establish if this model is still consistent with experimental constraints. The prime candidate for discovery is the CP-even Higgs, due to its relatively small mass. In the following we will see that most of the viable parameter space has already been explored by LEP and the LHC. All cross sections computed for our analysis were obtained from {\tt Madgraph 5}~\cite{Alwall:2014hca}, while branching ratios and the electroweak oblique parameters~\cite{Grimus:2008nb, Peskin:1991sw,Altarelli:1990zd} were computed with {\tt 2HDMC 1.8}~\cite{Eriksson:2009ws, Harlander:2013qxa}. We draw Feynman diagrams using {\tt TikZ-Feynman}~\cite{Ellis:2016jkw}.

\subsection{Experimental Constraints}\label{sec:exp}
In this Section we discuss current constraints and future probes of the new Higgs doublet, which are summarized in Fig.s~\ref{fig:intro_pheno} and~\ref{fig:CPeven_Heavy}. The parameter $\lambda_{345}$ is central to our discussion. It determines the maximum viable $m_H$ from Eq.~\eqref{eq:masses} and it sets $H$ couplings to SM fermions and bosons in Eq.~\eqref{eq:Hcouplings}. Lower bounds on $m_{H^\pm}^2 \sim -(\lambda_4+\lambda_5)v^2$ and $m_A=- \lambda_5 v^2$ determine a natural lower bound on $\lambda_{345}=\lambda_3+\lambda_4+\lambda_5$. In each panel of Fig.s~\ref{fig:intro_pheno} and~\ref{fig:CPeven_Heavy} we take $\lambda_{345}$ proportional to $m_{H^\pm}^2$. This explains the non-trivial dependence of $H$ phenomenology on $m_{H^\pm}$ in Fig.~\ref{fig:intro_pheno}. In Fig.~\ref{fig:CPeven_Heavy} we consider three different scenarios: $\lambda_{345}=(0.01,0.1,1)(2 m_{H^\pm}^2/v^2)$, corresponding to three levels of tuning: $1\%,10\%$ and no tuning. Tuning $\lambda_{345}$ small decreases $H$ couplings to fermions. This typically increases the allowed parameter space as shown in Fig.~\ref{fig:CPeven_Heavy}, but does not allow to decouple $H$ and can lead to $H\to \gamma\gamma$ become the dominant decay channel. 

There are areas of our parameter space that are theoretically inaccessible. These are shown in gray in Fig.s~\ref{fig:intro_pheno} and~\ref{fig:CPeven_Heavy}. At large $m_H$ and $\lambda_{345}$ there is no real solution for $\lambda_{1,2}$ that gives the observed SM Higgs mass. This happens when the argument of the square root in Eq.~\eqref{eq:masses} becomes negative. The second set of theoretical constraints arises from running of the quartics. At large $m_H$ and small $v_1$, $\lambda_1$ becomes large and one can get low energy Landau poles from $d\lambda_1/dt\simeq (3/4\pi^2)\lambda_1^2$. A similar situation occurs from the running of $\lambda_{4,5}$ when $m_A$ and/or $m_{H^\pm}$ become large, as shown in Fig.~\ref{fig:intro_pheno}. The remaining constraints in Fig.s~\ref{fig:intro_pheno} and~\ref{fig:CPeven_Heavy} are discussed in the next two Subsections, starting with direct searches.

\subsubsection{Direct Searches}

{\bf Charged Higgs} LEP-II gives a lower bound on the charged Higgs mass from $H^+H^-$ pair production~\cite{Abbiendi:2013hk}: $m_{H^\pm}\gtrsim 80\;{\rm GeV}$. This is shown in red in Fig.~\ref{fig:intro_pheno}. The constraint on $m_{H^\pm}$ comes from a combination of $\tau\nu$ and $cs$ final states and assumes the absence of $H^\pm \to W^\pm H$ decays. In the presence of a light neutral Higgs $m_H=12$~GeV the bound is slightly relaxed to $m_{H^\pm}\gtrsim 73\;{\rm GeV}$~\cite{Abbiendi:2013hk}. Note that the direct searches at LEP were performed for $m_{H^\pm}> 38$~GeV. Masses below $39.6$~GeV are excluded by measurements of the $Z$ boson width~\cite{ALEPH:2002aa} which receives a contribution from the charged Higgs~\cite{Abdallah:2003wd}. The LHC is mostly sensitive to $H^\pm$ production via top decays. Just like the $Z$, the top has a small width from electroweak interactions $\Gamma_t \simeq$~GeV, so the branching ratio $t\to H^+ b$ can be sizeable given that it is proportional to the top Yukawa coupling. We find that LHC searches for $t\to H^+ b$ with $H^+\to \tau^+\nu_\tau$ and $H^+\to c\bar s$~\cite{CMS:2016szv, CMS:2019sxh, CMS:2016qoa, CMS:2014kga} are sensitive to ${\rm BR}(t\to H^+ b)$ down to a few percent. The bound is shown in pink in Fig.s~\ref{fig:intro_pheno} and~\ref{fig:CPeven_Heavy}. When the mass difference between $H^\pm$ and $A$ becomes $\mathcal{O}(m_W)$, also the decays $H^\pm \to W^\pm A$ become relevant~\cite{PhysRevLett.123.131802}, but we do not consider this parameter space in our analysis since it is disfavored by bounds on $h\to AA$ and Electroweak precision measurements. For larger $m_{H^\pm}$, when top decays are not kinematically allowed, direct searches for $H^\pm$, that typically target $pp\to \bar t b H^+$ and decays to $tb$ and $\tau \nu$, are not yet sensitive to our parameter space~\cite{ATLAS:2020jqj, Sirunyan:2020hwv, Aaboud:2018cwk,CMS:1900zym,CMS:2019yat,Aaboud:2016dig,CMS:2018ect, CMS:2016szv, CMS:2014cdp}. To conclude this brief overview of the charged Higgs, it is interesting to notice that a CMS search for stau pair production~\cite{CMS:2019eln} has a sensitivity comparable to LHC searches targeting $H^+$ single production. The latter can be decoupled by making $v_1$ small, while pair production rates are fixed by gauge invariance. In the CMS search the staus decay to $\tau$'s plus a light neutralino ($m_\chi = $~GeV). The analysis is thus sensitive to $pp \to H^+ H^{-}\to \tau^- \tau^+ \nu_\tau \bar \nu_\tau$. Naively superimposing the cross section limit from this search with $H^+H^-$ VBF production gives a bound $m_{H^\pm} \gtrsim 150$~GeV. The DY pair production cross section is too small and does not give a constraint. This shows that the LHC can already set a (almost) model-independent bound on $H^\pm$ and warrants a more detailed analysis.

{\bf CP Even Higgs} LEP searches target mainly associated production $e^+e^-\to HZ$, which is controlled by the coupling $g_{HVV}$ in Eq.~\eqref{eq:Hcouplings}. The most sensitive channel is $H\to \bar b b$ for $m_{H}>2 m_b$~\cite{Schael:2006cr}, but even below this threshold LEP retains a comparable cross section sensitivity (down to $2 m_\mu$) by targeting different decays~\cite{Acciarri:1996um}. The strongest constraint above $2 m_b$ is set by the combined searches for $HZ$ production by the four LEP experiments~\cite{Schael:2006cr}. This is shown in yellow in Fig.s~\ref{fig:intro_pheno} and~\ref{fig:CPeven_Heavy}. The bound has a non-trivial dependence on $\lambda_{345}$. When $\lambda_{345} \ll \lambda_2$ the decays to SM fermions targeted by LEP are suppressed, but $HZ$ production is enhanced. In this limit our new CP even Higgs is fermiophobic and searches for $e^+e^-\to HZ\to \gamma\gamma Z$ become relevant. The LEP bound~\cite{ACHARD200228} dominates the light blue shaded area in Fig.~\ref{fig:CPeven_Heavy} for $m_H \lesssim 65$~GeV. When $\lambda_{345} \simeq \lambda_2$, the coupling to the $Z$ is as small as it can be: $g_{HVV}\simeq \mathcal{O}(v_1^3/v^3)$, but the decay to $\bar b b$ is enhanced compared to the limit $\lambda_{345} \ll \lambda_2$.  

To conclude the discussion of LEP constraints, it is interesting to notice that if we compare the model-independent VBF cross-section $e^+e^-\to H H \nu_e \nu_e$ with LEP pair production constraints~\cite{Schael:2006cr} we find sensitivity in the mass range $10\;{\rm GeV} \lesssim m_{H} \lesssim 25\;{\rm GeV}$. This is a bound that relies only on the electroweak doublet nature of $H_1$. It applies to a Higgs decaying mostly to $\bar b b$. LEP searches for $HH$ are optimized for CP-violating couplings, but the signal topology is not appreciably affected by the CP properties of the couplings~\cite{Schael:2006cr}. However this is a rough estimate of the actual constraint (given also that the search is not designed for VBF production) and it would be interesting to perform a dedicated collider study.

The main LHC constraints on $H$ arise from measurements of Higgs couplings, discussed in the next Section, and direct searches for $H\to\gamma\gamma$~\cite{Aad:2014ioa, Sirunyan:2018aui}. As noted above, when $\lambda_{345}\ll \lambda_2$ the CP-even Higgs couplings to fermions are suppressed and its BR to $\gamma\gamma$ becomes $\mathcal{O}(1)$. The LHC bounds on $H\to \gamma\gamma$ is shown in light blue in Fig.~\ref{fig:CPeven_Heavy}. 

Other direct searches for SM and BSM Higgses, $H\to\gamma\gamma$~\cite{ATLAS:2020tws, ATLAS:2020pvn, CMS:2020omd, Aaboud:2016tru, Aad:2014ioa, Sirunyan:2018aui, CMS:2014onr, Aad:2012yq, CMS:2012zwa, CMS:2015ocq, CMS:2017yta, Aaboud:2017yyg,Mariotti:2017vtv, Khachatryan:2015qba}, $H\to \tau^+\tau^-$~\cite{CMS:2020dvp, Aaboud:2017sjh, Aaboud:2016cre, Aad:2014vgg, Sirunyan:2018zut, CMS:utj, CMS:2013hja, CMS:2016pkt, CMS:2016rjp, CMS:2018klr, CMS:2017epy, CMS:2019pyn,CMS:2015mca, CMS:2018hqb}, $H\to \mu^+\mu^-$~\cite{Aad:2020xfq, Sirunyan:2020two, Aaboud:2019sgt, CMS:2019zrl, CMS:2016tgd, CMS:2016rme, Sirunyan:2019tkw}, $H\to \bar bb$~\cite{Aad:2020jym, Aad:2020eiv, Aad:2020vbr, ATLAS:2020syy, CMS:2020wwv, CMS:2020gxr, CMS:2018jfd, CMS:2013jda, Aad:2019zwb, CMS:2016ncz}, $H\to Z\gamma$~\cite{Aaboud:2017uhw}, $H\to W^+W^-, ZZ$~\cite{ATLAS:2020coj, Aad:2020mkp, ATLAS:2020qut, ATLAS:2020uti, CMS:2020dkv, CMS:bxa, CMS:2017vpy, Aad:2015agg, Aaboud:2017gsl, Aaboud:2017rel, Aaboud:2018eoy, CMS:2013pea, CMS:2013ada, CMS:2016jpd, CMS:2016noo} and $H\to ZA$~\cite{CMS:2019wml, CMS:2015mba, CMS:2019kjn, CMS:2016qxc}, give weaker constraints than the indirect probes discussed in the next Section and shown in Fig.~\ref{fig:CPeven_Heavy}. 

At lower masses, $0.3 \;{\rm GeV} \lesssim m_H\lesssim 5$~GeV, we have constraints from LHCb searches for rare $B$ meson decays~\cite{Aaij:2012vr}, from the CHARM beam dump experiment~\cite{BERGSMA1985458, Clarke:2013aya} and excess cooling of SN1987A~\cite{Krnjaic:2015mbs, 10.1143/ptp/84.2.233}. For $\lambda_{345}\gtrsim 10^{-2}$ both these constraints and the LHC bound on $\Gamma(h\to HH)$, discussed in the next Section, exclude the whole mass range. At lower values of $\lambda_{345}$ all the probes of a low mass Higgs, which are mainly sensitive to the coupling to fermions $g_{H\psi\psi}\sim \lambda_{345}/\lambda_2$, loose sensitivity. However proposals for future beam dump experiments~\cite{Berlin:2018pwi, na62, Alekhin:2015byh}, the HL-LHC projections for LHCb~\cite{Gligorov:2017nwh} and proposed long-lived particle experiments~\cite{Feng:2017vli, Gligorov:2017nwh, Evans:2017lvd, Chou:2016lxi} can cover most of the viable parameter space down to the lowest masses that we consider: $m_H \simeq 300$~MeV. 

{\bf CP Odd Higgs} In the mass range $0.3 \;{\rm GeV} \lesssim m_A\lesssim 5$~GeV, $A$ is excluded by LHCb searches for $B\to K^{(*)}\mu\mu$~\cite{Aaij:2012vr} and by the CHARM beam dump experiment~\cite{BERGSMA1985458, Clarke:2013aya}. The only exception are two mass windows ([2.95, 3.18]  GeV and [3.59, 3.77]  GeV) vetoed from the LHCb analysis to suppress the backgrounds from $J/\psi$ and $\psi^\prime$ production. Furthermore, $A$ can be lighter than $m_h/2$ only if $H$ is heavier than $m_h/2$ due to the indirect constraint on $\Gamma(h\to AA)$ and $\Gamma(h\to HH)$ discussed in the next Section. For $m_A<145$~GeV direct searches for $e^+e^-\to H_1 H_2$ at LEP~\cite{Schael:2006cr} exclude a large fraction of our parameter space due to the large $AH$ production cross section at the $Z$ pole,
\be
g_{ZAH}&\simeq&-\frac{g}{2 c_{\theta_W}}(p_A+p_H)\, .
\ee
This is shown in red in Fig.~\ref{fig:CPeven_Heavy}. We do not show this constraint in Fig.~\ref{fig:intro_pheno} since it completely overlaps with other bounds. LHC searches for BSM and SM Higgses, listed for the CP even Higgs, do not add new constraints to our parameter space.
The only exception are LHC searches for $pp \to A\to ZH$ and $pp\to A\to Zh$~\cite{Aad:2020ncx, Aad:2015wra, CMS:2013eua,CMS:2014yra, CMS:2018xvc}. For $m_A \gtrsim 220\;{\rm GeV}$ the LHC is already sensitive to an interesting range of our parameter space~\cite{CMS:2018xvc}. Most of this region is already excluded by the presence of low energy Landau poles, but a more detailed experimental study would be interesting.

Finally we can consider a light axion-like $A$, with $m_A \ll v$. However to be consistent with experiment this possibility requires $v_1 \ll \Lambda_{\rm QCD}$, to suppress the couplings of $A$ to the SM. This forces also $H$ to be light, at odds with LHC bounds on ${\rm BR}(h\to HH)$ discussed in the next Section.

\subsubsection{Indirect Constraints}\label{sec:indirect}
Measurements of low energy flavor changing processes, such as $B\to X_s \gamma$ and $B \to \tau \nu$, are powerful probes of our charged Higgs~\cite{Barbieri:1993av, Gori:2017tvg}. In both cases the charged Higgs contributes at the same order as the leading SM diagram, given by a $W$ boson exchange. If we include other well-measured processes sensitive to the charged Higgs ($B\to K^*\gamma$, $B_s\to \mu^+\mu^-$, $D_s\to \tau\nu$, $B\to K^{(*)} l^+ l^-$, $R_D$, $R_{D^*}$ and $B_s \to \phi \mu^+\mu^-$)~\cite{Arbey:2017gmh}, we find the bound in light green in Fig.~\ref{fig:intro_pheno} and in pink in Fig.~\ref{fig:CPeven_Heavy}.

The second set of indirect constraints that we need to consider arises from LHC measurements of SM Higgs couplings. At small $v_1$ and fixed masses they read 
\be
\frac{g_{hVV}-g_{hVV}^{\rm SM}}{g_{hVV}^{\rm SM}}\simeq -\frac{v_1^2}{2v^2}\left(1-\frac{\lambda_{345} v^2}{m_h^2-m_H^2}\right)^2\, , \quad \quad
\frac{g_{h\psi\psi}-g_{h\psi\psi}^{\rm SM}}{g_{h\psi\psi}^{\rm SM}}\simeq -\frac{v_1^2}{2v^2}\left(1-\frac{\lambda_{345}^2 v^4}{\left(m_h^2-m_H^2\right)^2}\right)\, , \label{eq:SMHiggsMasses}
\ee
so $v_1\lesssim v$ insures that couplings to both vector bosons and fermions are consistent with experiment.  As $m_H$ approaches the SM Higgs mass at fixed $\lambda_{345}$, the sensitivity to $v_1/v$ increases. We also have regions where $\lambda_{345} v^2 \simeq m_h^2-m_H^2$ and most sensitivity is lost. In Fig.~\ref{fig:CPeven_Heavy} we show in purple projections for the HL-LHC and a future lepton collider at 1$\sigma$. We take projected sensitivities from Table 1 in~\cite{deBlas:2019wgy}\footnote{This choice is conservative as it allows the presence of additional new physics that modifies Higgs couplings. In our model alone, we could have used a more constrained fit with two universal coupling modifiers for fermions and bosons and a free width to new particles.}. To represent future lepton colliders we use ILC with unpolarized beams and $\sqrt{s}=250$~GeV. 
 Current bounds from the LHC~\cite{CMS:2020gsy, ATLAS:2020qdt} are not shown in the Figure. They give the bound $v_1/v\lesssim 0.45\div0.55$ and completely overlap with other constraints. The only exception are new decay modes of the SM Higgs. 
The new CP-even Higgs $H$ is lighter than the SM Higgs. So for small enough $m_H$ we also have to consider the new decay width $\Gamma(h\to H H)$. Direct searches for decays to four SM particles via two intermediate states, $h\to HH \to 4 {\rm SM}$~\cite{Aad:2020rtv, CMS:2020bni, Sirunyan:2020eum, Sirunyan:2018pzn, Sirunyan:2018mbx, Aaboud:2018esj}, are less constraining than the indirect bound set by the dilution of SM branching ratios. LHC measurements of Higgs couplings give an upper bound on $\lambda_{345}$ if $m_H < m_h/2$, from Eq.~\eqref{eq:trilinear}. If we consider the latest ATLAS combination of Higgs couplings measurements~\cite{ATLAS:2020qdt}, we have a global signal strength $\mu=1.06\pm0.07$. From CMS~\cite{CMS:2020gsy} we have $\mu=1.02^{+0.07}_{-0.06}$. A very rough combination, assuming uncorrelated Gaussian errors, gives a $2\sigma$ error $\delta \mu_{95\%}\simeq 0.1$.  This implies ${\rm BR}(h\to HH)\lesssim 0.1$ at $2\sigma$ and hence $\lambda_{345}\lesssim 10^{-2}$. If we restrict ourselves to the Yukawa couplings in Eq.~\eqref{eq:Yukawas} there is not much that we can do to relax this constraint. For example, increasing $g_{h b b}$ to compensate for the new decay mode decreases $g_{hVV}$, as shown in Eq.~\eqref{eq:SMHiggsMasses}. In more generality, barring detailed constructions that exploit flat directions in Higgs couplings constraints, we need $\lambda_{345}= \mathcal{O}(10^{-2})$ for $m_H<m_h/2$, given our current knowledge of Higgs couplings. This is shown in Fig.~\ref{fig:CPeven_Heavy}. When we tune $\lambda_{345}$ to be small we have no constraint, while in the more natural parameter space $m_H < m_h/2$ is excluded. Note that if we take $\lambda_{345}$ small the same reasoning leads to $m_A > m_h/2$, since $\lambda_{hAA}\simeq (\lambda_{345}-2\lambda_5)v$.

To conclude, ElectroWeak Precision Tests~\cite{ALEPH:2005ab}, mainly the $S$, $T$ and $U$ oblique parameters~\cite{Peskin:1991sw,Altarelli:1990zd} constrain mostly the mass difference between $A$ and $H^\pm$ that breaks the custodial symmetry. The bound is displayed in light blue in Fig.~\ref{fig:intro_pheno}. A more detailed analytical discussion of oblique parameters, custodial symmetry and CP in 2HDMs can be found in~\cite{Haber:2010bw, Pomarol:1993mu}.

\section{Weak Scale Triggering Low Energy Landscape for Small CC}\label{sec:landscape}

We now present a vacuum selection mechanism in the context of the landscape, where the weak scale as a trigger is crucially needed to find a vacuum with tiny enough cosmological constant. We imagine there is a ``UV landscape" containing moderately many vacua, not enough to find vacua with our small Cosmological Constant (CC). The UV landscape scans the CC and the Higgs mass(es) without scanning dimensionless couplings~\cite{ArkaniHamed:2005yv}.

We also imagine a separate ``IR landscape", with $n_\phi$ ultra-light, weakly coupled scalars $\phi_i$, each with a (spontaneously broken) $Z_2$ discrete symmetry, potentially giving a factor of $2^{n_\phi}$ more vacua.  The $\phi_i$ also couple to a trigger operator ${\cal O}_T$. If $\langle {\cal O} _T \rangle$ is too small, the $2^{n_\phi}$ vacua of the $\phi_i$ sector are all degenerate and they don't help with making smaller vacuum energies possible. If $\langle {\cal O}_T \rangle$ is too big, the symmetry is broken so badly that only one vacuum remains for each $\phi_i$, and there is again no way to find small vacuum energy. The only way to find small vacuum energy is to tune Higgs vacuum expectation values so that $\mu_S^{\Delta_T} < \langle {\cal O}_T \rangle < \mu_B^{\Delta_T}$. Thus using the weak scale as a trigger allows us to tie solutions to the cosmological constant and hierarchy problems. 

Our low-energy effective theory, contains in addition to the SM or the type-0 2HDM, a ``IR landscape" consisting of $n_\phi$ scalars $\phi_i$. In first approximation the scalars are uncoupled, and each have a $Z_2$ discrete symmetry, described by the potential:
\be
V_{N\phi} =\sum_{i=1}^{n_\phi} \frac{\epsilon_i^2}{4}\left(\phi_i^2-M_{*,i}^2\right)^2\, . \label{eq:scalars}
\ee
In absence of new symmetries or dynamics below $M_*$, we take the $\phi_i$ vevs $M_{*,i}$ to be the fundamental scale of the theory $\mathcal{O}(M_*)$. $\epsilon_i$ is an order parameter that quantifies the breaking of the shift symmetry on $\phi_i$, such that $m_{\phi_i} \sim \epsilon_i M_{*,i} \ll M_*$ is technically natural. 
We assume that the cosmological constant and the Higgs mass(es) squared are scanned uniformly in a ``UV landscape", which has $N_{\rm UV}$ vacua, with $N_{\rm UV}$ too small to find a vacuum with small enough CC. The smallest CC in the UV landscape is $\simeq M_*^4/N_{\rm UV}$. In vacua where Higgs mass(es) squared are $\sim v^2$ the minimal CC is larger and we call this value of the CC $\Lambda_*$.

\begin{figure*}[!t]
\begin{center}
\includegraphics[width=0.48\textwidth]{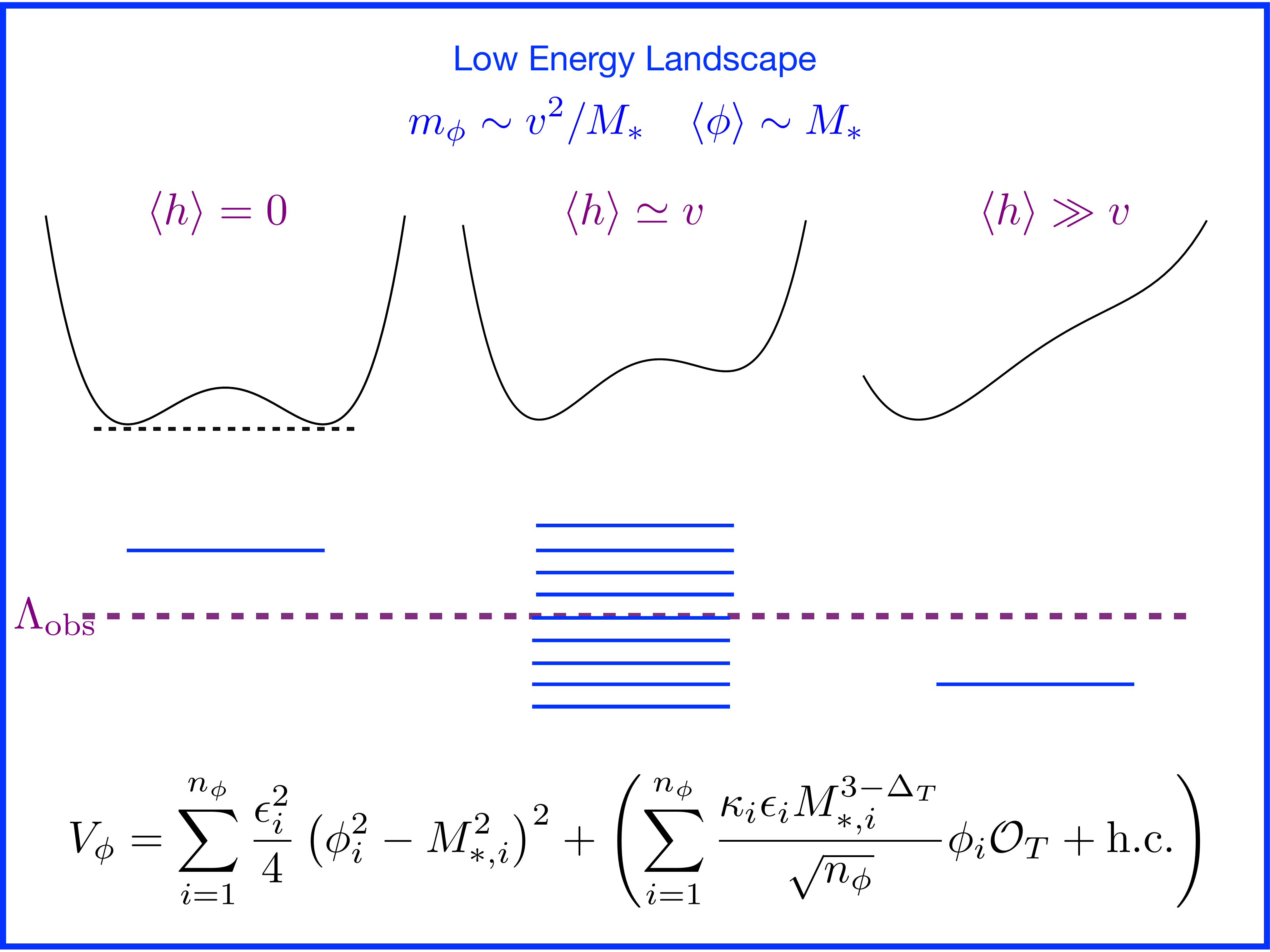}
\end{center}
\vspace{-.3cm}
\caption{The landscape contains a UV sector and an IR sector (in this Figure). The high energy sector is generated by fields of mass close to the cutoff $m_\Phi\sim M_*$ and does not have enough vacua to scan the CC down to $\Lambda_{\rm obs}\simeq {\rm meV^4}$, but can scan the Higgs mass(es) $m_H^2$ down to the weak scale. The low energy sector is generated by fields of mass $m_\phi \sim v^2/M_*$ and has a number of non-degenerate minima dependent on the Higgs vev. When $\langle h \rangle \simeq v$ we can scan the CC down to its observed value.}
\label{fig:landscape}
\end{figure*}

\begin{figure*}[!t]
\begin{center}
\includegraphics[width=0.7\textwidth]{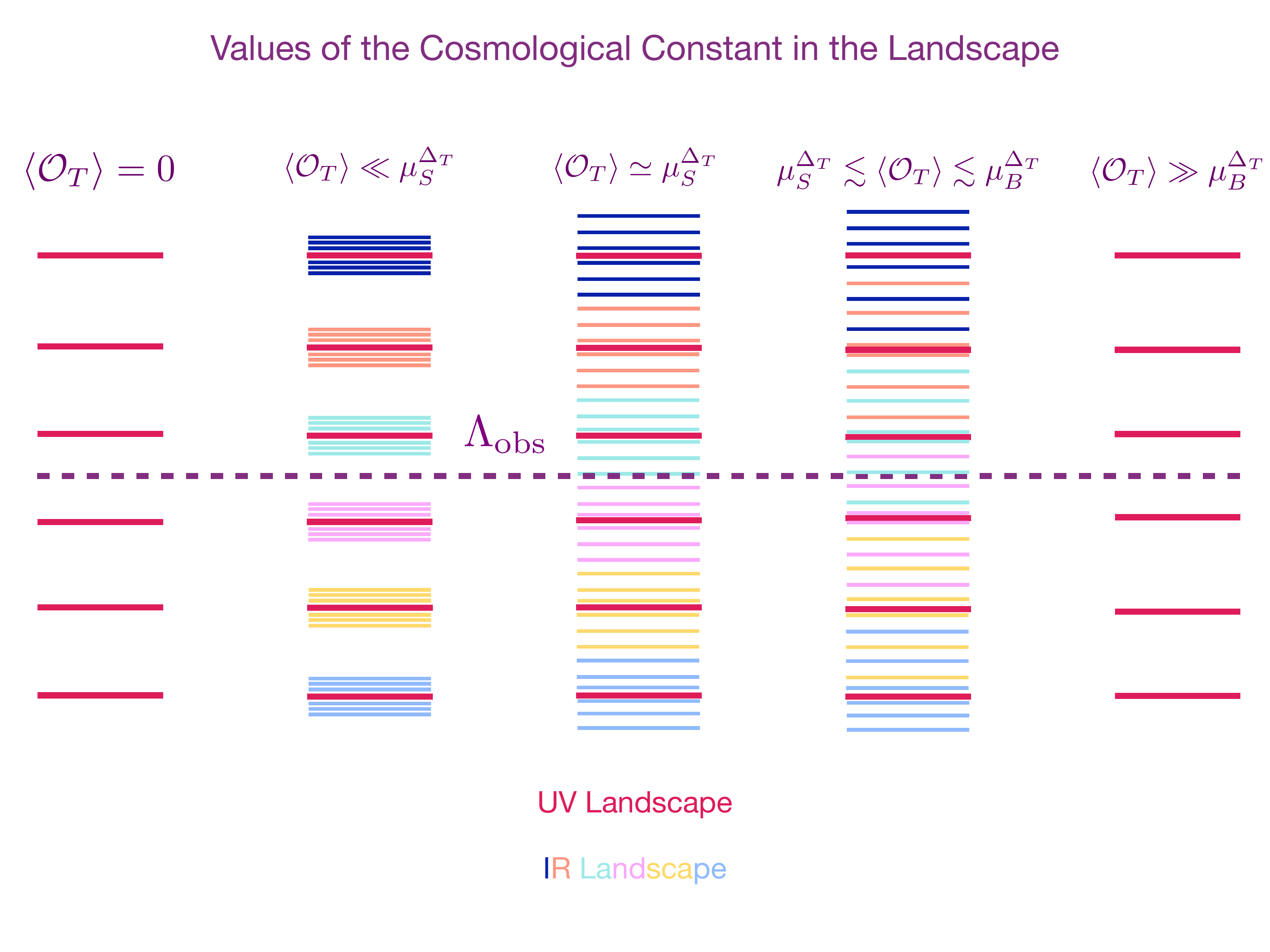}
\end{center}
\vspace{-.3cm}
\caption{Values of the Cosmological Constant in our two-sectors landscape. When $\langle \mathcal{O}_T \rangle=0$ the UV landscape does not have enough minima to scan the CC from $M_*^4$ down to $\Lambda_{\rm obs}\simeq {\rm meV^4}$. The minimal value of the CC in the landscape is $M_*^4/N_{\rm UV} \gg {\rm meV^4}$. When $\langle \mathcal{O}_T \rangle \neq 0$ the degeneracy in the vacua of the low energy landscape in Fig.~\ref{fig:landscape} is broken and if $\mu_S^{\Delta_T}\lesssim \langle \mathcal{O}_T \rangle\lesssim \mu_B^{\Delta_T}$ we can harness the full potential of its $2^{n_\phi}$ vacua and scan the CC down to ${\rm meV^4}$. If $\langle \mathcal{O}_T \rangle \gg \mu_B^{\Delta_T}$ the low energy landscape loses all its minima but one and the minimal CC in the landscape is again $M_*^4/N_{\rm UV} \gg {\rm meV^4}$.}
\label{fig:landscape_2}
\end{figure*}

We now imagine that each of the $\phi_i$ also couples to our weak scale trigger operator ${\cal O}_T$, 
\be
V_{N\phi T}=\sum_{i=1}^{n_\phi} \frac{\kappa_i \epsilon_i M_{*,i}^{3-\Delta_T}}{\sqrt{n_\phi}} \phi_i \mathcal{O}_T+{\rm h.c.} 
\label{eq:VphiT}
\ee
Here $\kappa_i$ parametrizes an additional weak coupling, breaking the $(Z_2)^{n_\phi}$ symmetry down to a single diagonal $Z_2$. Note that gravity loops also couple the different sectors, but the coupling to gravity doesn't break the $(Z_2)^{n_\phi}$ discrete symmetry that we have when $\kappa \to 0$, (and at any rate, induces parametrically minuscule cross-quartics of order $\epsilon_i^2 \epsilon_j^2 \phi_i^2 \phi_j^2$). 
The structure of our IR landscape is depicted in Fig.~\ref{fig:landscape} while their role in scanning the CC is sketched in Fig.~\ref{fig:landscape_2}. The interaction in Eq.~\eqref{eq:VphiT} makes the number of minima in the landscape sensitive to the value of $\langle \mathcal{O}_T\rangle$. If\footnote{For simplicity we have dropped the subscript $i$, assuming that all $\epsilon_i$, $\kappa_i$ and $M_{*,i}$ are close to a common value.}
\be
\langle \mathcal{O}_T \rangle \gtrsim \frac{\epsilon}{\kappa} \sqrt{n_\phi} M_{*}^{\Delta_T} \equiv \mu_B^{\Delta_T}\, ,
\ee
some minima are lost, as shown in Fig.~\ref{fig:landscape}, which makes it impossible for the CC to have the observed value.

If $\langle \mathcal{O}_T\rangle$ is too small 
\be
\langle \mathcal{O}_T\rangle \lesssim  \frac{\sqrt{n_\phi}}{\epsilon \kappa}\frac{\Lambda_*}{M_*^4} M_*^{\Delta_T} \equiv \mu_S^{\Delta_T}
\ee
the degeneracy between the minima of Eq.~\eqref{eq:scalars} is not lifted enough to scan the CC down to $({\rm meV})^4$. This defines the two scales $\mu_S$ and $\mu_B$. To see how these two opposite pressures on the vev of $\langle \mathcal{O}_T\rangle$ select the weak scale we need to specify the field content of $\mathcal{O}_T$. In the two following Sections we discuss $\mathcal{O}_T=G \widetilde G$ and $\mathcal{O}_T=H_1 H_2$.

\subsection{SM Trigger of the Landscape}\label{sec:SMtrigger}
The simplest trigger $\mathcal{O}_T$ is already present in the SM. It is given by the familiar $G\widetilde G$ operator introduced in Section~\ref{sec:trigger} that we now couple to the $n_\phi$ scalars in the low energy landscape,
\be
V_{N\phi G}&=& \frac{1}{32\pi^2}\sum_{i=1}^{n_\phi} \left(\frac{\phi_i}{f_i}+\theta\right)G \widetilde G\, , \quad G \widetilde G\equiv \epsilon^{\mu\nu\rho\sigma}\sum_a G_{\mu\nu}^a G_{\rho\sigma}^a\, .
\label{eq:trigger}
\ee
Here we only briefly discuss how to use this trigger in the context of our landscapes. We give more details in the next Section for the $H_1 H_2$ trigger.

In the notation of the previous section we have
\be
\mathcal{O}_T=G \widetilde G\, , \quad \frac{1}{f_i} = \frac{32 \pi^2 \kappa_i \epsilon_i}{\sqrt{n_\phi}M_{*, i}}\, .
\ee
We imagine that one of the usual mechanisms solves the strong CP problem, leaving at low energy a residual $\theta$ angle smaller than $10^{-10}$. We also impose $\langle \phi_i\rangle /f_i \lesssim 10^{-10}/n_\phi$ to avoid re-introducing the problem. 
To study the effect of $V_{N\phi G}$ in Eq.~\eqref{eq:trigger} we can move $\phi_i$ into the quark mass matrix with an anomalous chiral rotation and use chiral perturbation theory to get
\be
V_{N\phi G}&\simeq& \left\{\begin{array}{c}f_\pi^2(\vev) m_\pi^2(\vev) \left(\sum_i\frac{\phi_i}{f_i}+\theta\right)^2+...\, ,\quad \langle h \rangle \lesssim \frac{\Lambda_{\rm QCD}(\vev)}{y_u}\, , \\ \Lambda_{\rm QCD}^4(\langle h \rangle)\left(\sum_i\frac{\phi_i}{f_i}+\theta\right)^2+... \, ,\quad \langle h \rangle \gtrsim \frac{\Lambda_{\rm QCD}(\vev)}{y_u}\, .\end{array}\right. \label{eq:VHphi}
\ee

In the previous equation we have introduced $\Lambda_{\rm QCD}(v_*)$, which is the chiral condensate with quark masses proportional to the vev $v_*$. Similarly $f_\pi^2(v_*)$ and $m_\pi^2(v_*)$ are the values of these parameters with EW symmetry breaking at the scale $v_*$. Note that the dependence on $v_*$ saturates when $\Lambda_{\rm QCD}(v_*) \geq v_*$ and QCD itself becomes the main source of EW symmetry breaking. 

The potential in Eq.~\eqref{eq:VHphi} makes the number of minima in the landscape sensitive to the value of the Higgs vev $\vev$. When $\vev$ is too large some minima are lost, when it is too small the minima of Eq.~\eqref{eq:scalars} remain almost degenerate. To see why minima are lost when $\langle h \rangle$ is large consider the limit $\Lambda_{\rm QCD}^4 M_*/f \gg \epsilon^2 M_*^4$, where Eq.~\eqref{eq:VHphi} dominates over the potential in~\eqref{eq:scalars}. Then at the minimum Eq.~\eqref{eq:VHphi} is effectively fixing $\sum_i (\phi_i/f_i)=-\theta$. We can implement this condition as a Lagrange multiplier
\be
\mathcal{L}=\lambda\left(\sum_i\frac{\phi_i}{f_i}+\theta\right)-V_{N\phi}\, .
\ee
From the point of view of this Lagrangian fixing $\sum_i (\phi_i/f_i)=-\theta$ corresponds to $\lambda/f_i \gg \partial V_{N\phi}/\partial \phi_i$, so when we try to solve the cubic equation
\be
f_i \frac{\partial V_{N\phi}}{\partial \phi_i}=f_i \epsilon_i^2 \phi_i\left(\phi_i^2-M_{*,i}^2\right)=\lambda\, ,
\ee
we are guaranteed to find at most one solution. This would be true also if $V_{N\phi}$ was a periodic potential. This discussion shows that in the limit of large $\langle h \rangle$ all minima in the low energy landscape (but one) are lost.

In summary we have an upper and a lower bound on $\vev$ that depend on the mass of the scalars in the landscape and on $\Lambda_*$ (the smallest CC in the UV landscape). If we imagine that both opposing ``pressures" are saturated at the same value of $\vev$, then in the multiverse this is the only value consistent with Weinberg's anthropic argument. We have measured this value to be the weak scale $v$, so the mass scale in the low energy landscape and the residual CC must be: 
\be
m_\phi \simeq \frac{f_\pi m_\pi}{\min[f, \sqrt{f M_*/\theta}]}\, , \quad \Lambda_* \simeq \left(N_2 f_\pi m_\pi \sqrt{\frac{\theta M_*}{f}}\right)^2\lesssim (100\; {\rm keV})^4 \left(\frac{\theta}{10^{-10}}\right)\, ,
\label{eq:mechanism}
\ee 
where $m_\pi$ and $f_\pi$ are those observed in our universe. In Eq.~\eqref{eq:mechanism} the value of $m_\phi$ determines whether minima are lost or not, so it depends on the term that dominates the QCD potential of the new scalars. This can either be the linear one if $\theta > M_* /f_i$ or the quadratic one in the opposite limit. On the contrary $\Lambda_*$ is sensitive only to the linear term, since $\Lambda_{\rm QCD}^4\theta \phi/f$ provides the only difference between the value of the potential at the two minima $\phi \simeq \pm M_*$. This shows that we cannot have an axion lighter than the $\phi_i$'s and the strong CP problem has to be solved at higher energies. 

There is a priori no reason why the two pressures on $\langle h \rangle$ are saturated at the same scale, given that they arise from two distinct physical requirements ($V_{N\phi G}\lesssim V_{N\phi}$ and $\Lambda_* \simeq V_{N\phi G}^{\rm min}$). In general we expect a range around the weak scale, $\mu_S \lesssim \langle h \rangle \lesssim \mu_B$, to be viable. We further expand on this point in the next Section. 

\subsection{Type-0 2HDM Triggering of the Landscape}\label{sec:2HDMtrigger}
We now consider the case where the triggering operator is ${\cal O}_T={\cal O}_H=H_1 H_2$, so that we have 
\bea
V^{(I)} & = &  \sum_{i=1}^{n_\phi}\left[\frac{\epsilon^2}{4}\left(\phi_i^2-M_{*}^2\right)^2 + \frac{\epsilon \kappa}{\sqrt{n_\phi}} M_* \phi_i H_1 H_2\right] + V_H^{(I)}\, .
\label{eq:Vphi}
\eea
For simplicity we have dropped the subscript $i$, assuming that all $\epsilon_i$, $\kappa_i$ and $M_{*,i}$ are close to a common value. The Higgs potential reads
\bea
V_H^{(I)} & = & (m_1^2)^{(I)} |H_1|^2 + (m_2^2)^{(I)} |H_2|^2 + \Lambda^{(I)}+ \nn \\
&+&\frac{\lambda_1}{2}|H_1|^4+\frac{\lambda_2}{2}|H_2|^4 
+\lambda_3 |H_1|^2|H_2|^2 +\lambda_4 |H_1 H_2|^2 
+\left(\frac{\lambda_5}{2}(H_1 H_2)^2 +{\rm h.c.}\right) \nn \\
&+& Y_u q H_2 u^c +Y_d q H_2^\dagger d^c + Y_e l H_2^\dagger  e^c\, .
\eea
Here $I = 1, \cdots, N_{\rm UV}$ labels vacua in the UV landscape. We imagine that $\Lambda^{(I)}$, $(m_{1,2}^2)^{(I)}$ are uniformly distributed between $(-M_*^4, M_*^4)$, $(-\Lambda_H^2, \Lambda_H^2)$ where $\Lambda_H$ is the Higgs cutoff. Of course the simplest choice is to assume $\Lambda_H^2 \sim M_*^2$, but it is possible to have $\Lambda_H^2 \ll M_*^2$ and on occasion we will consider $\Lambda_H$ much smaller than $M_*$. 

The first term in Eq.~\eqref{eq:Vphi} has a large $(Z_2)^{n_{\phi}}$ discrete symmetry. The couplings $\phi_i H_1 H_2$ break it down to a single diagonal $Z_2$ under which all $\phi_i$'s are odd and $H_1 H_2$ is odd. The small parameter $\epsilon$ is a measure of the small breaking of the shift symmetry $\phi_i \to \phi_i + c_i$,  while $\kappa$ is a further weak coupling of $\phi_i$ to $H_1 H_2$, which breaks $(Z_2)^{n_{\phi}}$ down to the diagonal $Z_2$. By spurion analysis we should also include 
\begin{equation}
\Delta V_{ij} \sim \sum_{i,j} \epsilon^2 \kappa^2 \phi_i \phi_j M_*^2 \label{eq:loop_scanning}
\end{equation}
in the potential for $\phi$, which is logarithmically induced by a one-loop diagram. Now, suppose we are in a region of the big landscape where the operator $(H_1 H_2)$ is {\it not} triggered, i.e. $\mu^2 \equiv \langle H_1 H_2 \rangle = 0$, say with $m_{1,2}^2 > 0$ and close to the cutoff $\Lambda_H^2$. From the UV landscape, we have a distribution of vacua with CC splittings of order $\Delta \Lambda_{\rm UV}\simeq M_*^4/N_{\rm UV}$.
When $\kappa = 0$, each of these vacua is $2^{n_\phi}$ degenerate, as in the first column of Fig.~\ref{fig:landscape_2}. Turning on $\kappa$, from $\Delta V_{ij}$ in Eq.~\eqref{eq:loop_scanning} we get CC splittings of order $\epsilon^2 \kappa^2 M_*^4$. If this splitting was much bigger than $\Delta \Lambda_{\rm UV} \simeq M_*^4/N_{\rm UV}$, we would already finely scan the CC. So, we assume that $\kappa$ is small enough so this splitting is much smaller than the splitting in the UV landscape,
\bea
\epsilon^2 \kappa^2 M_*^4 & \ll & \frac{M_*^4}{N_{\rm UV}}. \label{eq:k_splitting}
\label{eq:1}
\eea

Note that if we tune down the Higgs masses squared to $m_1^2$ and $m_2^2$, the CC splitting in the UV landscape increases as
\bea
\Delta \Lambda_{\rm UV} (m_1^2,m_2^2) & \sim & \frac{M_*^4}{N_{\rm UV}} \frac{\Lambda_H^2}{|m_1^2|} \frac{\Lambda_H^2}{|m_2^2|},
\eea
so if the condition in Eq.~\eqref{eq:k_splitting} is satisfied, then obviously the loop-induced $\epsilon^2 \kappa^2 M_*^4$ splitting gets even smaller relative to $\Delta \Lambda_{\rm UV}$.
Thus we must have $\mu^2 =\langle H_1 H_2 \rangle \neq 0$ in order to be able to find a vacuum with the CC much smaller than $\Delta \Lambda_{\rm UV} \simeq M_*^4/N_{\rm UV}$.

Now let's look at the region in the landscape with $m_{1,2}^2 < 0$, and look at tree-level where $\mu^2 = \sqrt{|m_1^2| | m_2^2|}$.
If $\mu^2$ is too big, we tilt the $\phi_i$ potentials so much as to lose one of the vacua, as shown in Fig.~\ref{fig:landscape}. This happens for $\mu^2 \gtrsim \mu_B^2$ where $\mu_B^2$ is determined from 
\begin{equation}
\epsilon^2 M_*^4 \sim \kappa \epsilon M_*^2 \mu_B^2 \to \mu_B^2 \sim \frac{\epsilon}{\kappa} M_*^2\, .
\end{equation}
When $\mu$ drops below $\mu_B$, we want the splittings in the $N_{\rm IR} = 2^{n_{\phi}}$ vacua, now of order $\epsilon \kappa \mu^2_B M_*^2 \simeq \kappa^2 \mu_B^4$, to be much larger than 
\bea
\Delta \Lambda_{\rm UV} (m_1^2, m_2^2) & \sim & \frac{M_*^4}{N_{\rm UV}} \frac{\Lambda_H^4}{|m_1^2 m_2^2 |} \sim \frac{M_*^4}{N_{\rm UV}} \frac{\Lambda_H^4}{\mu_B^4}.
\eea
So we should have
\bea
\kappa^2 & \gg &  \frac{M_*^4}{N_{\rm UV}} \frac{\Lambda_H^4}{\mu_B^8}.
\label{eq:2}
\eea
Putting Eq.s~\eqref{eq:1} and~\eqref{eq:2} together, we have
\bea
\frac{\Lambda_H^2 M_*^2}{N_{\rm UV}^{1/2} \mu_B^4} & \ll & \kappa \ll \frac{M_*}{\mu_B} \frac{1}{N_{\rm UV}^{1/4}}.
\eea
This forces $N_{\rm UV} \gg (\Lambda_H^2 M_*/\mu_B^3)^4$. So for $\mu_B^2 \simeq v^2$, suppose we take the simplest possibility where $\Lambda_H \sim M_* \sim M_{\rm Pl}$.
Then, the above inequality tells us that $N_{\rm UV} \gg 10^{180}$. In this case, our mechanism is clearly irrelevant. There would be more than enough vacua in the UV landscape to simply tune down one Higgs and the CC. For our mechanism to be relevant, we would like to have $N_{\rm UV} \ll (M_*^4/\Lambda_{\rm obs})(\Lambda_H^2/v^2) \simeq10^{120} \Lambda_H^2/v^2$.
Thus we must have
\bea
\frac{\Lambda_H^8 M_*^4}{v^{12}} & \ll & 10^{120} \frac{\Lambda_H^2}{v^2},
\eea
and we get an upper bound on the Higgs cutoff: $\Lambda_H \ll 10^{12} \ {\rm GeV}$.
Note that since $N_{\rm UV} \gg (\Lambda_H^2 M_*/\mu_B^3)^4$, and $\kappa \ll \frac{M_*}{\mu_B} \frac{1}{N_{\rm UV}^{1/4}}$, we have also an upper bound on the coupling of the new scalars to the Higgses
\bea
\kappa \ll \frac{\mu_B^2}{\Lambda_H^2} \sim \frac{v^2}{\Lambda_H^2}.
\eea
With these conditions, $\Lambda_H \ll 10^{12} \ {\rm GeV}$ and $\kappa \ll v^2/\Lambda_H^2$, satisfied, our mechanism works. For instance if we take $\Lambda_H \simeq 10^6$~GeV, $M_*\simeq M_{\rm GUT}\simeq 10^{16}$~GeV, $\kappa\simeq 10^{-5}$, $\epsilon\simeq 10^{-30}$, $N_{\rm UV}\simeq 10^{77}$, we have $\mu_S\simeq 100$~GeV, $\mu_B\simeq 6$~TeV and the smallest CC in the UV landscape corresponds to an Hubble size of $\mathcal{O}(10 R_\odot)$. With this choice of parameters we have $n_\phi \simeq 120$ light scalars with mass $m_\phi\simeq 10^{-5}$~eV in the IR landscape which scan the CC down to its observed value. 
In the next Section we discuss $\phi_i$ dark matter, but let us mention here two cosmological constraints necessary for our mechanism to work. First, we must have $H_{\rm inf} \ll M_*$ during inflation. This ensures that the fluctuations of the $\phi_i$ during inflation are small, so that after inflation, our Hubble patch has the $\phi_i$s in the basin of attraction of one of the $N_{\rm IR} = 2^{n_{\phi}}$ minima. Obviously the condition $H_{\rm inf} \ll M_*$ is trivially satisfied for $M_* \sim M_{\rm Pl}$. We also want the $\phi_i$s to be massive enough to actually oscillate and reach their minima. Minimally we should have $m_{\phi_i} \gg H_{\rm today}$. Putting $m_{\phi} \sim \epsilon M_* \sim  \kappa\mu_B^2/M_*$, we have that $\kappa v^2/M_* \gg H_{\rm today}$ which gives a lower bound on $\kappa$.
\bea
\frac{v^2 M_*}{M_{\rm Pl}^3} \ll \kappa \ll \frac{v^2}{\Lambda_H^2}.
\eea
The resulting condition on $\Lambda_H$, $\Lambda_H^2 M_* \ll M_{\rm Pl}^3$, is trivially satisfied once $\Lambda_H \le M_* \le M_{\rm Pl}$.
Indeed, $m_\phi \simeq H_{\rm today}$ is the limit in which $\Delta \Lambda_{\rm UV} < H_{\rm today}^2 M_*^2 \le H_{\rm today}^2 M_{\rm Pl}^2 \sim \Lambda_{\rm obs}$ and the maximum scan of $\phi$ is smaller than the observed CC and thus our mechanism would be irrelevant. 

As we keep dropping $\mu$, at some point the splitting $\epsilon \kappa \mu^2 M_*^2 \sim \kappa^2 \mu^2 \mu_B^2$ will eventually become smaller than $\Delta \Lambda_{\rm UV}$. This happens for $\mu = \mu_S$, where $\mu_S$ is defined by
\bea
\kappa^2 \mu_S^2 \mu_B^2 & \sim & \frac{M_*^4}{N_{\rm UV}} \frac{\Lambda_H^4}{\mu_S^4}, 
\eea
which determines $\mu_S$ as
\bea
\mu_S & \sim & \frac{M_*^{2/3} \Lambda_H^{2/3}}{\mu_B^{1/3}} \frac{1}{N_{\rm UV}^{1/6} \kappa^{1/3}}.
\eea
Below $\mu \sim \mu_S$, the extra scanning of $N_{\rm IR} = 2^{n_{\phi}}$ vacua cannot bring the smallest CC down, and the minimum CC shoots back up to $M_*^4/N_{\rm UV}$. A schematic plot of the smallest CC in the landscape, as a function of $\mu^2$, is shown in Fig. 7. As we have seen, only for $\mu_S^2 \lesssim  \mu^2 \lesssim \mu_B^2$ can the power of the extra $2^{n_{\phi}}$ vacua be harnessed to exponentially suppress the CC's we can get from the landscape. 

\begin{figure*}[!t]
\begin{center}
\includegraphics[width=0.55\textwidth]{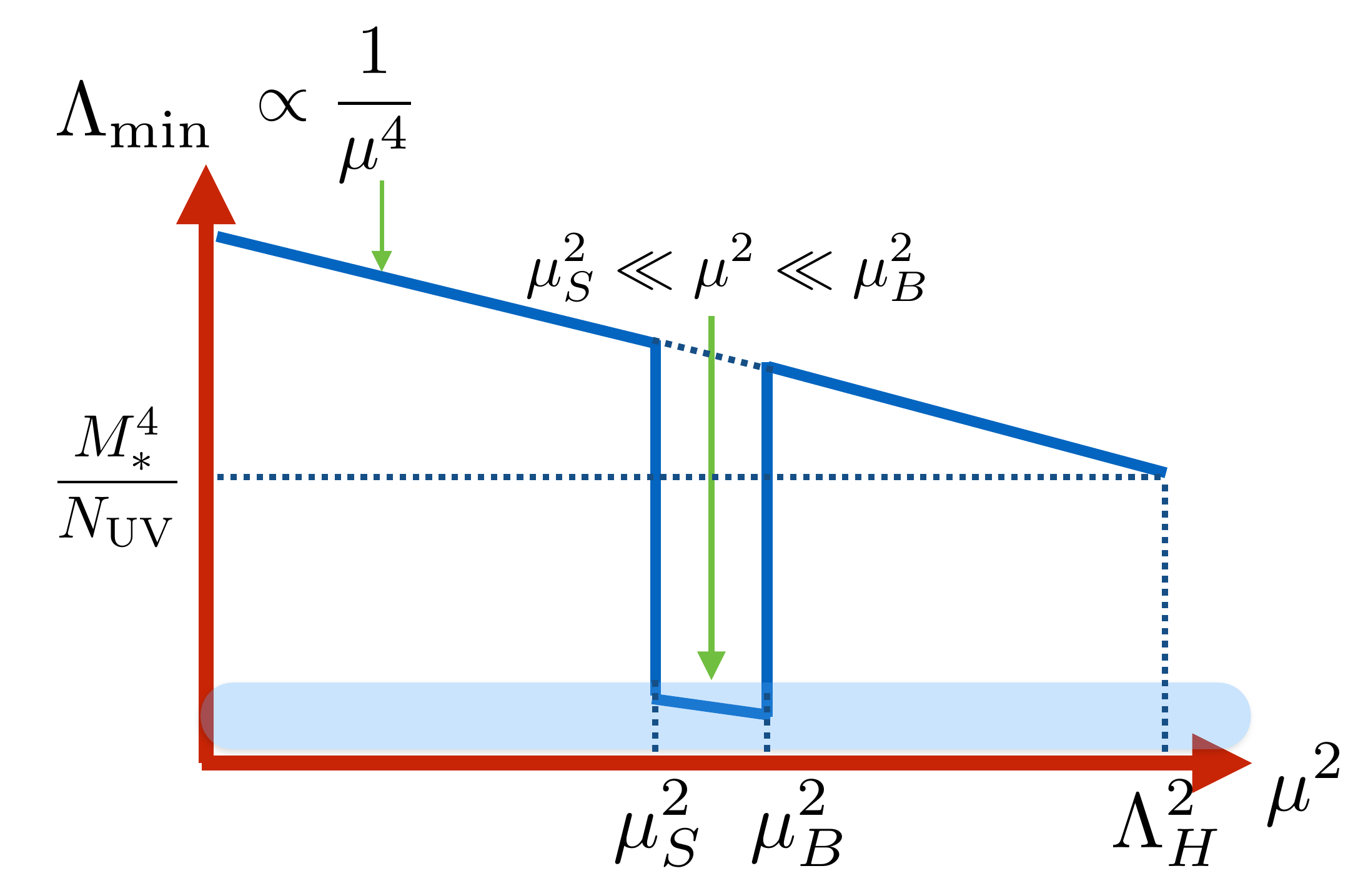}
\end{center}
\vspace{-.3cm}
\caption{The smallest CC in the landscape as a function of $\mu^2\equiv \langle H_1 H_2 \rangle$. In the light blue area the CC is smaller than its observed value, while for $\mu^2>\mu_B^2$ or $\mu_S<\mu_S^2$ it is much larger, $M_*^4/N_{\rm UV}\gg {\rm meV}^4$.}
\label{fig:hdkim}
\end{figure*}
We can perform a similar analysis for the case where $m_1^2 < 0$ and $m_2^2 > 0$, in which $\mu^2 \sim \sqrt{|m_1^2|} \Lambda_{\rm QCD}^3/m_2^2$. Here keeping $\mu^2 < \mu_S^2$ is easy since $\mu^2$ is naturally tiny ($\simeq \Lambda_{\rm QCD}^3/\Lambda_H$). Instead the constraint is in making $\mu$ big enough for the splitting $\kappa^2 \mu^2 \mu_B^2$ to be bigger than $\Delta \Lambda_{\rm UV}$. Clearly $\Delta \Lambda_{\rm UV}(\mu)$ is minimized when $m_1^2 \sim -\Lambda_H^2$. Then $\mu^2 \sim \Lambda_H \Lambda_{\rm QCD}^3/m_2^2$ and we obtain
\bea
\Delta \Lambda_{\rm UV}(-\Lambda_H^2, m_2^2) & \sim & \frac{M_*^4}{N_{\rm UV}} \frac{\Lambda_H^2}{m_2^2} \sim \frac{M_*^4}{N_{\rm UV}} \frac{\mu^2 \Lambda_H}{\Lambda_{\rm QCD}^3}.
\eea
To scan the CC to its observed value, we need the splittings in the IR landscape to be larger than $\Delta \Lambda_{\rm UV}$. Then we must have
\bea
\kappa^2 \mu^2 \mu_B^2 & \gg & \frac{M_*^4}{N_{\rm UV}} \frac{\mu^2 \Lambda_H}{\Lambda_{\rm QCD}^3},
\eea
which gives a lower bound on $\mu_B^2$,
\bea
\mu_B^2 & \gg & \frac{1}{N_{\rm UV} \kappa^2} \frac{M_*^4 \Lambda_H}{\Lambda_{\rm QCD}^3}.
\eea
If this happens we can also find small CC vacua in this part of the landscape. But note that we can never find a vacuum that looks like our world here. While the W/Z bosons are massive, near the cutoff $\Lambda_H$, the fermions are massless in the effective field theory beneath $\Lambda_H$. If we integrate out $H_2$, the 4 fermi operators $(q q^c) ( e e^c)/m_2^2$ are generated and leptons also get minuscule masses $\sim \Lambda_{\rm QCD}^3/m_2^2 \sim \mu^2/\Lambda_H$ after chiral symmetry breaking. But if we suppose the parameters of the model are such as to have $\mu^2 \lsim \mu_B^2 \lsim v^2$, the lepton masses are suppressed by at least by a factor of $v/\Lambda_H$ compared to our world. In this situation for atoms to form, the temperature of the universe must drop by a factor of $v/\Lambda_H$ further relative to our universe, meaning that the CC must be further smaller by a factor of $(\frac{v}{\Lambda_H})^4$ before atoms can form. It could easily be that $N_{\rm IR} = 2^{n_{\phi}}$ is not large enough to realize this possibility. Thus while finding vacua with tiny CC suppressed by $1/N_{\rm IR} = 2^{-n_{\phi}}$ is possible with $m_1^2 \sim - \Lambda_H^2$, forcing $m_2^2 > 0$ to be tuned small, these worlds look nothing like ours. It is only possible to get a world that looks like ours with $m_1^2 <0$ and $m_2^2 < 0$.
As we have seen in our discussion of the phenomenology of this model, since the weak scale is set by the largest of the Higgs VEVs, this forces the existence of new light charged and neutral Higgs states which we cannot decouple or tune away.

\section{Ultralight Dark Matter from Weak Scale Triggers}\label{sec:dark_matter}
In this Section we describe a very interesting feature of our low energy landscape: it provides new dark matter (DM) candidates whose relic abundance is rather insensitive to the high energy history of our universe and it is only determined by the DM mass and its coupling to the SM. 

Take $H_1 H_2$ as a trigger. At the time of the electroweak phase transition (EWPT), $\langle H_1 H_2 \rangle$ turns on, displacing the new scalars by an amount $\Delta \phi=\mathcal{O}(M_*)$. The corresponding energy density $\rho_\phi \sim m_\phi^2 M_*^2 \sim \kappa^{2} v^4$ depends only on the $\phi$'s coupling to the SM $\kappa$. So to a first approximation the relic density today depends only on $\kappa$ and the scalar mass $m_\phi$. This is reminiscent of WIMPs, whose abundance is uniquely determined by their coupling to the SM and their mass. In the case of WIMPs initial conditions are washed out by electroweak interactions with the SM bath, in our case by the EWPT displacement triggered by $H_1 H_2$. WIMPs are insensitive to initial conditions if the Universe is reheated not too far below the dark matter mass, while in our case we need the initial SM temperature to be above that of the EWPT. The above statements can be made more explicit by computing the relic density of the scalars from their classical equation of motion
\be
\ddot \phi + 3 H \dot \phi + \frac{\partial V_{N\phi}}{\partial \phi}+ \frac{\kappa \epsilon M_*}{\sqrt{n_\phi}} \langle H_1 H_2\rangle_T =0\, .
\label{eq:cosmo}
\ee

At temperatures $T \gg v$ we have $\langle H_1 H_2\rangle_T=0$, so initially we can neglect the last term in the equation. After the EWPT, the interaction with the Higgs only gives a constant shift to the scalar potential. Therefore the only effect of the coupling to the Higgs is to give a kick $\Delta \phi=\mathcal{O}(M_*)$ to the new scalars at $T\simeq v$.

The second and third terms in the equation determine when the new scalars start to oscillate, transitioning from dark energy to dark matter. Since the quartic and trilinear coupling of $\phi$ are largely subdominant at the scale of the mass we have $\partial V_{N\phi}/\partial \phi \simeq m_\phi^2 \phi$. So the evolution of $\phi$ is determined by the value of $m_\phi$ in units of Hubble. We imagine that the $\phi$ potential in Eq.~\eqref{eq:scalars} has its zero-temperature form throughout the history of the Universe, i.e. the sector generating this potential has dynamics above its reheating temperature to avoid domain-wall problems. Then $m_\phi$ is temperature independent in our analysis.
There are two relevant regimes for $m_\phi$. It can be larger or smaller than Hubble at the electroweak phase transition, $H(v)$. There are also two natural possibilities for the initial (i.e. $T\gg v$) displacement of the scalars from their minimum, either $\Delta \phi \sim M_*$ or $\Delta \phi < M_*$. We start by considering the case $m_\phi < H(v)$. The scalars are frozen in place by Hubble friction until after the EWPT. When the phase transition happens, the scalar potential is shifted by the Higgs vev, generating a displacement of $\mathcal{O}(M_*)$. Therefore, regardless of the initial misalignment, the new scalars start to oscillate and redshift as cold dark matter with $\Delta \phi \sim \mathcal{O}(M_*)$ when $m_\phi \simeq H$. In this case initial conditions change the relic density at most at $\mathcal{O}(1)$.

In the second case, $m_\phi > H(v)$, the scalars start to oscillate before the EWPT. Their initial energy density starts to redshift at $T>v$ and when $T \simeq v$ it is already smaller than $m_\phi^2 M_*^2$. Therefore the kick imparted by the EWPT is responsible for the dominant contribution to the energy density. 

\begin{figure*}[!t]
\begin{center}
\includegraphics[width=0.45\textwidth]{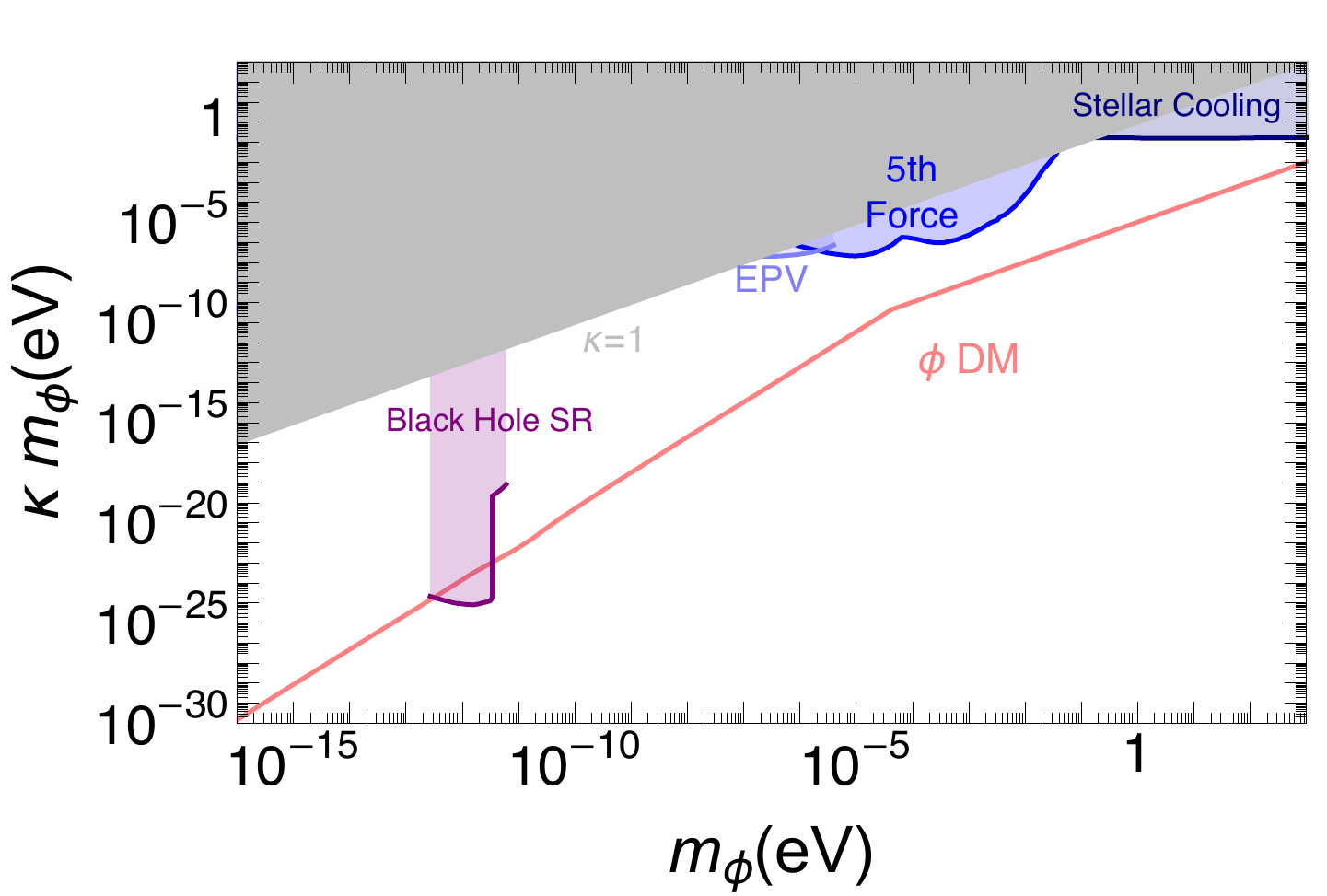}
\end{center}
\vspace{-.3cm}
\caption{Laboratory and astrophysical constraints on scalars coupled to the Higgs boson via the trilinear interaction $\kappa m_\phi \sum_{i=1}^{n_\phi}\phi_i |H|^2/\sqrt{n_\phi}$ (we neglect unimportant $\mathcal{O}(1)$ factors introduced by the mixing of the two Higgses). The bounds include tests of the equivalence principle~\cite{Smith:1999cr, Schlamminger:2007ht, Wagner:2012ui, Berge:2017ovy}, tests of the Newtonian and Casimir potentials (5th force)~\cite{Spero:1980zz, Hoskins:1985tn, Chiaverini:2002cb, Hoyle:2004cw, Smullin:2005iv, Kapner:2006si,Bordag:2001qi, Bordag:2009zzd,Turyshev:2006gm}, stellar cooling~\cite{Hardy:2016kme} and black hole superradiance~\cite{Arvanitaki:2014wva, Baryakhtar:2020gao}. The pink solid line shows the target given by the scalars being dark matter. We shaded in gray the region where $\kappa >1$ (i.e. $\Lambda_H \lesssim$~TeV).}
\label{fig:lablight}
\end{figure*}

Solving Eq.~\eqref{eq:cosmo} we obtain that the right relic density is given by
\be
\kappa &\sim& \frac{m_\phi^{3/4} M_{\rm Pl}^{1/4}}{v}\, , \quad H(v)> m_\phi \nn \\ 
\kappa &\sim&\sqrt{\frac{v}{M_{\rm Pl}}}\, , \quad H(v)\leq m_\phi \, .
\label{eq:coupling}
\ee
To highlight the parametrics we have used the rough approximation $T_{\rm eq}\sim v^2/M_{\rm Pl}$ for the temperature of matter radiation equality. We do not use this approximation in Figures and when quoting numerical results. Given that $\kappa$ determines the coupling of the new scalars to the SM, we have a target for ultralight dark matter and fifth force searches, shown in Fig.~\ref{fig:lablight}. In this Section for simplicity we neglect the $\mathcal{O}(1)$ difference between $v^2$ and $v_1 v_2$. We also take $\lambda_{345}\ll \lambda_2$ so that at leading order in $v_1/v$ we can neglect $\mathcal{O}(1)$ factors introduced in Fig.~\ref{fig:lablight} by the mixing of the two Higgses. The viable ranges for the dark matter mass and dark matter coupling are
\be
10^{-22}\;{\rm eV}\lesssim m_\phi \lesssim {\rm keV}  \, \quad 10^{-19}\lesssim \kappa \lesssim 10^{-5}\, . \, \label{eq:parameters}
\ee
The upper bound on the dark matter mass is determined by not lowering the cutoff $M_*$ below 10 TeV. Note that even at the upper end of this range $m_\phi \simeq $~keV the lifetime of the scalars is about $10^{20}$ times the age of the Universe. The lower bound is determined by astrophysical measurements of small scale structure. The precise lower bound on the DM mass is still the subject of active research, see for instance~\cite{Hui:2016ltb, Leong:2018opi, Irsic:2017yje, Armengaud:2017nkf, Bozek:2014uqa,Safarzadeh:2019sre, Bar:2018acw}. 

We have obtained our previous results neglecting $\phi^3$ and $\phi^4$ terms in the potential. These anharmonic terms are too small to have a measurable impact on structure formation. One conservative way to see this is to show that the effective Jeans length that they induce~\cite{Turner:1983he, Arvanitaki:2014faa} is smaller than the typical size of a galaxy ($\sim$~{\rm Mpc}) at all times between today and matter radiation equality. This is satisfied in all our DM parameter space. Imposing the same requirement on the Jeans length induced by the $\phi$ mass: $L_J(m_\phi)\sim (M_{\rm Pl}^2/\rho_\phi m_\phi)^{1/2}$ leads to $m_\phi \gtrsim 6\times 10^{-21}$~eV. This is consistent with observational bounds on the lightest viable DM mass and comparable to our theoretical lower bound in Eq.~\eqref{eq:parameters}. As mentioned above, establishing a precise lower bound on the DM mass is still the subject of active research and goes beyond the scope of this work. 

In Fig.~\ref{fig:lablight} we also show laboratory and astrophysical constraints on $\phi$ DM. They include tests of the equivalence principle~\cite{Smith:1999cr, Schlamminger:2007ht, Wagner:2012ui, Berge:2017ovy}, tests of the Newtonian and Casimir potentials (5th force)~\cite{Spero:1980zz, Hoskins:1985tn, Chiaverini:2002cb, Hoyle:2004cw, Smullin:2005iv, Kapner:2006si,Bordag:2001qi, Bordag:2009zzd, Turyshev:2006gm}, stellar cooling~\cite{Hardy:2016kme} and black hole superradiance~\cite{Arvanitaki:2014wva, Baryakhtar:2020gao}. Fifth force and equivalence principle constraints were translated on bounds on the trilinear coupling of a scalar coupled to the Higgs boson in~\cite{Graham:2015ifn,Piazza:2010ye}. The bound from superradiance is cut off by the quartic $\epsilon^2$ that at fixed $m_\phi$ and $\kappa$ scales as $\epsilon^2\sim m_\phi^4/(v^4 \kappa^2)$.

Future laboratory probes of our scalars include torsion balance experiments~\cite{Graham:2015ifn}, atom interferometry~\cite{Arvanitaki:2016fyj}, optical/optical clock comparisons and nuclear/optical clock comparisons~\cite{Arvanitaki:2014faa} and resonant mass detectors (DUAL and SiDUAL~\cite{Leaci:2008zza}). We do not show them in the Figure because they are $\mathcal{O}(15)$ orders of magnitude away from the $\phi$ dark matter line.

In addition to the laboratory and astrophysical constraints shown in the Figure, Planck's measurement of the power spectrum of isocurvature perturbations~\cite{Ade:2015lrj} sets a mild constraint on Hubble during inflation
$H_k \lesssim  10^{-5}  N_2 M_* (\Omega_c/\Omega_{\phi})\lesssim 10^{18}\;{\rm GeV} \; (\Omega_c/\Omega_{\phi})$.
The subscript $k$ means that Hubble is evaluated when the perturbation leaves the horizon $k=a H$ and is subsequently frozen. In quoting the bound we have used the most constraining scale measured by Planck $k_0=0.002$~Mpc$^{-1}$ and assumed isocurvature perturbations that are completely uncorrelated with curvature perturbations. 

As we noted at the end of Section~\ref{sec:trigger} we need to break the $H_1\to -H_1$ symmetry to avoid a domain wall problem. At the EW phase transition the new scalars spontaneously break the $Z_2$ providing an effective $B\mu$ large enough to avoid domain wall domination: $B\mu\simeq \kappa \epsilon M_*^2 \simeq \kappa^2 v^2 \gg v^4/M_{\rm Pl}^2$.

To conclude, note that also in the case of a $G \widetilde G$ trigger we can have $\phi$ dark matter. The relic density is set as in the case discussed above, but this time the new scalars get their $\mathcal{O}(M_*)$ kick at the QCD phase transition. The energy density at $T\simeq \Lambda_{\rm QCD}$ is $\sim m_\phi^2 M_*^2 \simeq f_\pi m_\pi M_*^2/\min[f^2, f M_*/\theta]$ and is determined by the ratio of $f$ and $M_*$. We find that the only viable parameter space is $m_\phi \lesssim 10^{-17}$~eV. Having $\phi$ dark matter for larger $m_\phi$ requires $M_* > 10^{-10} f/N_2$ at odds with the strong CP problem.  

{\emph{Acknowledgments.} ---} We thank A. Hook, M. Low, L.T. Wang for collaboration at an initial stage of this work. We thank A. Arvanitaki, T. Banks, M. Baryakhtar, S. Dimopoulos, M. Geller, M. Sher, S. Stefanelli, for useful discussions and comments. HDK is supported by the NRF of Korea grant, No. 2017R1A2B201074914.


\bibliography{refs}
\bibliographystyle{utphys}


\end{document}